\documentclass[preprint,12pt]{elsarticle}

\usepackage{amssymb}
\usepackage{graphicx}	% Include figure files
\usepackage{xspace}     % Include xspace
\usepackage[usenames,dvipsnames]{color}%mjt-this uses familiar color names
\newcommand{\Npart}{\mbox{$N_{\rm part}$}\xspace}
\newcommand{\Ncoll}{\mbox{$N_{\rm coll}$}\xspace}
\newcommand{\Nch}{\mbox{$N_{\rm ch}$}\xspace}
\newcommand{\Nqp}{\mbox{$N_{\rm qp}$}\xspace}
\newcommand{\Et}{\mbox{${\rm E}_T$}\xspace}

\newcommand{\sqsn}{\mbox{$\sqrt{s_{_{NN}}}$}\xspace}
\def\lsim{\raise0.3ex\hbox{$<$\kern-0.75em\raise-1.1ex\hbox{$\sim$}}}
\def\gsim{\raise0.3ex\hbox{$>$\kern-0.75em\raise-1.1ex\hbox{$\sim$}}}
\def\mean#1{\left<#1\right>}
\journal{Annals of Physics}

\begin{document}

\begin{frontmatter}



\title{Reminiscences of Wit Busza and 41 Years of p+A Physics}


\author{Michael J. Tannenbaum~\fnref{1}}

\address{Brookhaven National Laboratory, Upton, NY 11973-5000, USA}
\fntext[1] {Research supported by U.S. Department of Energy, DE-AC02-98CH10886.}
\begin{abstract}
One of the more memorable (and easiest) proposal to deal with when I served on Bob Wilson's Program Advisory Committee at NAL (Now Fermilab) from 1972-75 was Proposal-178, ``A study of the average multiplicity and multiplicity distributions in  hadron-nucleus collisions at high energies'', with only 4 authors, Wit Busza, Jerry Friedman, Henry Kendall and Larry Rosenson, as presented at the PAC meeting by Wit. What I remember was that he discussed only ONE 5 inch photomultiplier with a Cherenkov radiator in the beam to make this measurement of production of charged particles with angles up to 30 degrees in various nuclei, 40 hours requested. This turned out to be a ``seminal'' experiment leading to the Wounded Nucleon and other participant models. Subsequent $p(d)+A$ experiments from the AGS to RHIC, as well as alpha-alpha measurements at the CERN-ISR, will be discussed together with the various `participants' that they revealed.
\end{abstract}

\begin{keyword}


\end{keyword}

\end{frontmatter}

% \linenumbers

\section{E-178}
\label{sec:E178}
\begin{figure}[!h] 
     \centering
      \includegraphics[width=0.65\linewidth]{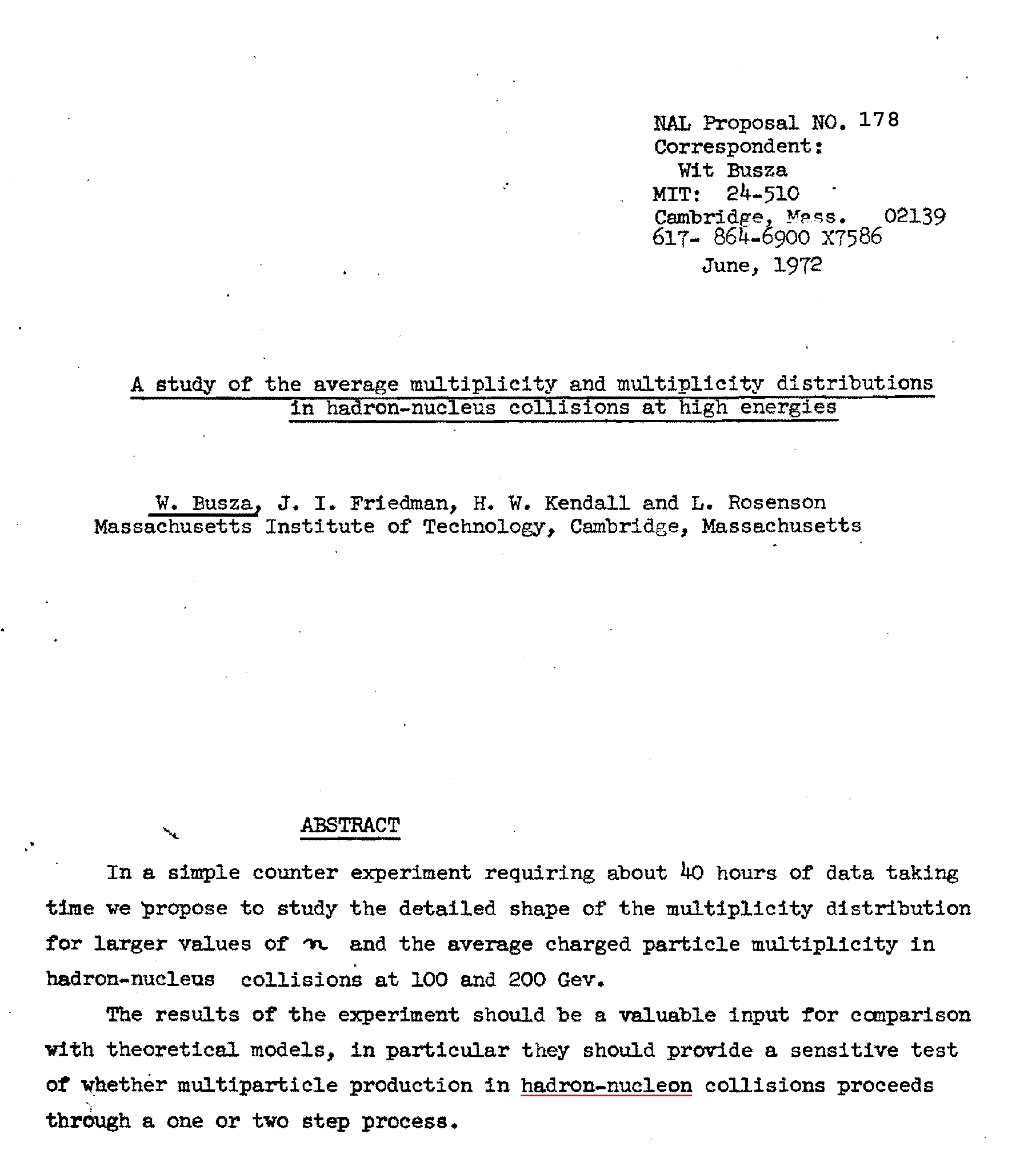} 
      \caption[]{Cover page of NAL Proposal NO. 178}
      \label{fig:Prop1}
   \end{figure}

\begin{figure}[!h] 
      \centering
      \includegraphics[width=0.7\linewidth]{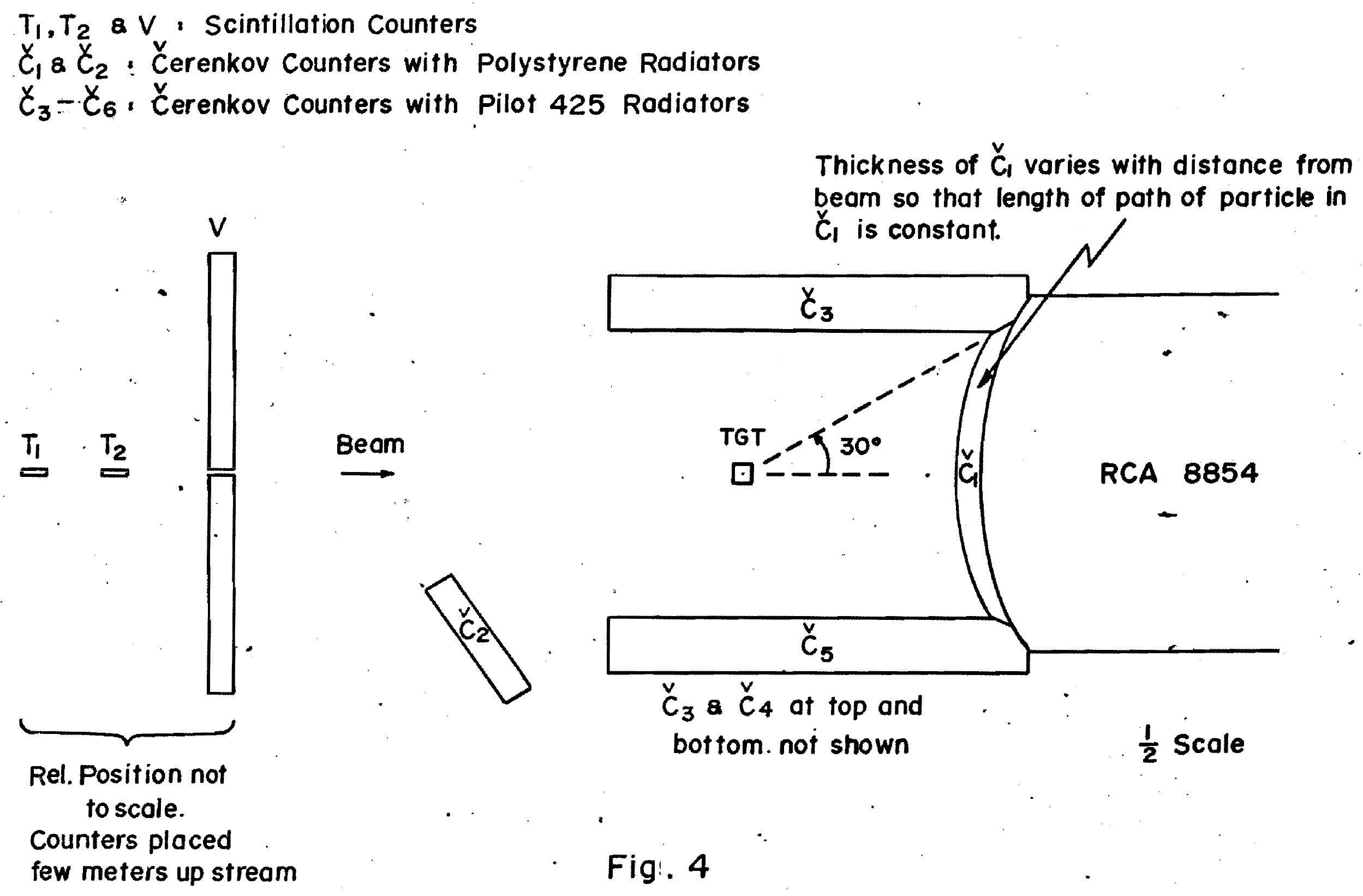} 
      \caption[]{Layout of P-178, dominated by the 5-inch diameter RCA 8854 photomultiplier.}
      \label{fig:PropLayout}
   \end{figure}

     The first experiment specifically designed to measure the dependence of the charged particle multiplicity in high energy  p+A collisions as a function of the nuclear size was performed by Wit Busza and collaborators (Fig.~\ref{fig:Prop1}) at Fermilab using beams of $\sim 50-200$ GeV/c hadrons colliding with various fixed nuclear targets. 
The experiment was unique in the small (but distinguished) number of authors; and in its size, which would literally fit on a table top (Fig.~\ref{fig:PropLayout}).

\begin{figure}[!h] 
      \centering
      \includegraphics[width=0.65\textwidth,angle=0.2]{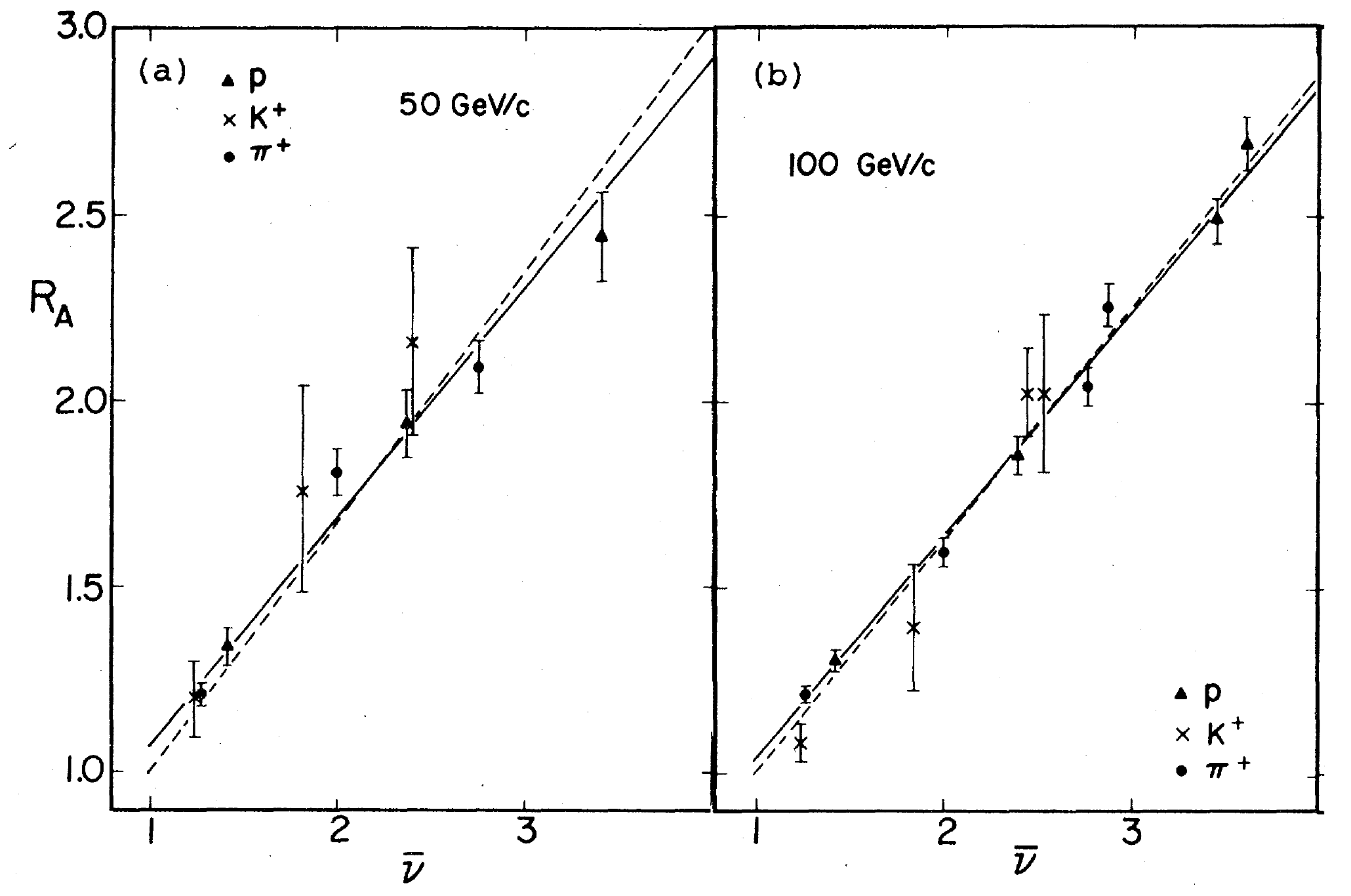}\hspace*{-0.5pc}
            \includegraphics[width=0.39\textwidth,angle=-1]{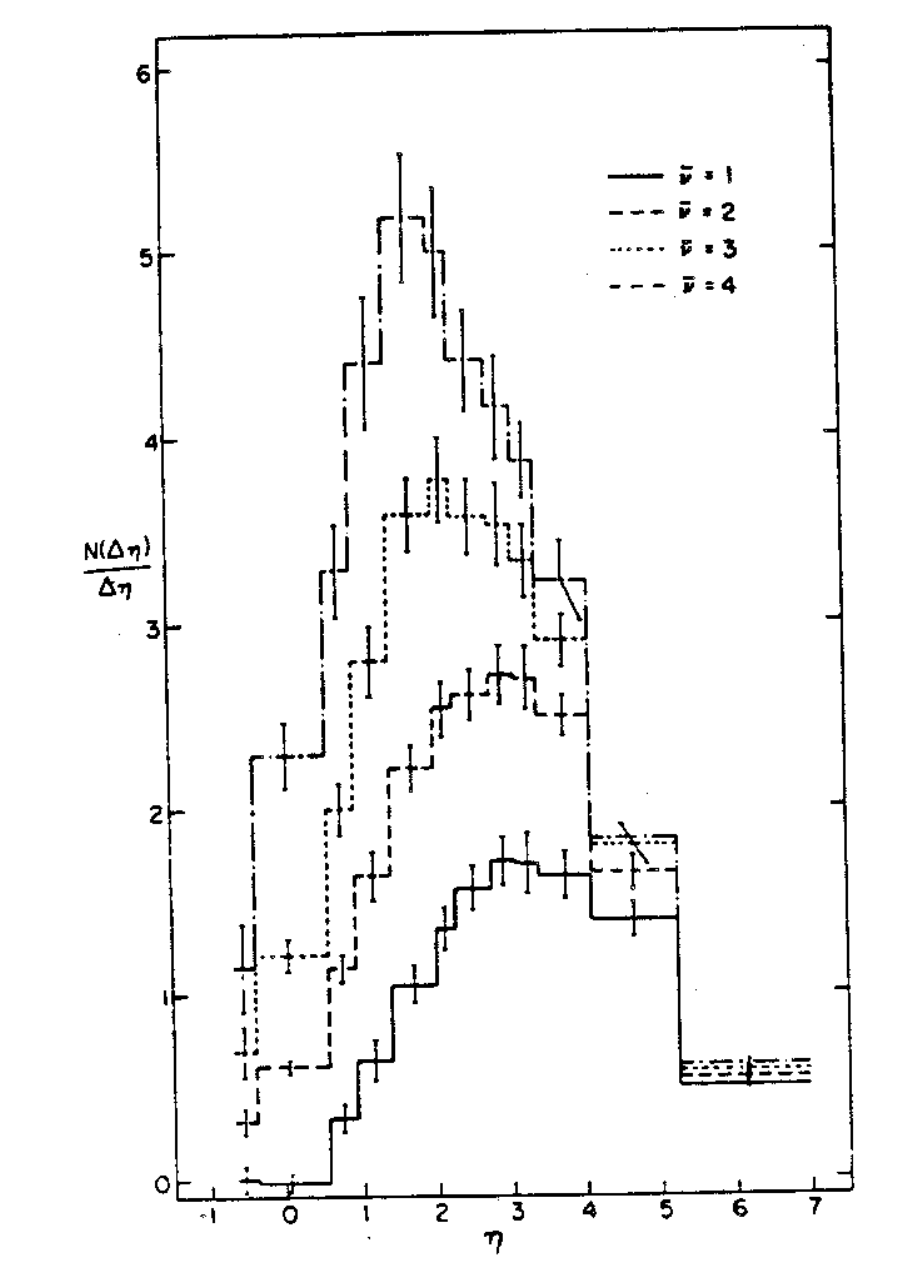}
      \caption[]{a)(left) $R_A=\mean{\Nch}_{hA}/\mean{\Nch}_{hp}$ as a function of the average thickness of each nucleus given in terms of the mean free path, $\overline{\nu}=\mean{\Ncoll}$~\cite{EliasPRD22} for 50 and 100 GeV/c h+A  collisions; b) (right) Charged particle multiplicity density, $d\Nch/d\eta$, as a function of $A$ (represented by $\overline{\nu}$) for 200 GeV/c p+A collisions~\cite{HalliwellPRL39}.}
      \label{fig:pAdists}
      \end{figure}

         They found the extraordinary result~\cite{BuszaPRL34} that the average charged particle multiplicity $\mean{\Nch}_{hA}$ in hadron+nucleus (h+A) interactions was not simply proportional to the number of collisions (absorption-mean-free-paths), $\Ncoll=\nu$, but increased much more slowly, proportional to the number of participants \Npart . Thus, relative to h+p collisions (Fig.~\ref{fig:pAdists}a)~\cite{EliasPRD22}: 
         \begin{equation}
         R_A=\mean{\Nch}_{hA}/\mean{\Nch}_{hp}=(1+\overline{\nu})/2 \label{eq:npartscaling}\qquad.
         \end{equation} 
Since the different projectiles, $h=\pi^+, K^+, p$ in Fig.~\ref{fig:pAdists}a have different mean free paths, the fit to the same straight line in terms of $\overline{\nu}$ is convincing.

      The other striking observation (Fig.~\ref{fig:pAdists}b)~\cite{HalliwellPRL39} was that a relativistic incident proton could pass through e.g. $\nu=4$ absorption-mean-free-paths of a target nucleus and emerge from the other side; and furthermore there was no intra-nuclear cascade of produced particles 
(a stark difference from what would happen to the same proton in a macroscopic 4 mean-free-path hadron calorimeter). %%%
In the forward fragmentation region of 200 GeV/c p+A collisions, within 1 unit of rapidity from the beam, $y^{\rm beam}=6.0$, there was essentially no change in $d\Nch/d\eta$ as a function of $A$, while at mid-rapidity ($y^{\rm cm}_{_{NN}}\sim 3.0$), $d\Nch/d\eta$ increased with $A$ together with a small backward shift of the peak of the distribution  resulting in a huge relative increase of multiplicity in the target fragmentation region, $\eta<1$ in the laboratory system. 
These striking features of the $\sim 200$ GeV/c fixed target hadron-nucleus data ($\sqsn\sim 19.4$ GeV) showed the importance of taking into account the time and distance scales of the soft multi-particle production process including quantum mechanical effects \cite{FishbaneTrefilPRD9,FishbaneTrefilPLB51,GottfriedPRL32,ASGoldhPRD7,BialasCzyzPLB51,BoIngvarNPB88}. 
\section{The Wounded Nucleon Model} 
The observations in Fig.~\ref{fig:pAdists} had clearly shown that the target nucleus was rather transparent so that a relativistic incident nucleon 
could make many successive collisions while passing through the nucleus, and emerge intact. %%%
Immediately after a  relativistic nucleon interacts inside a nucleus, the only thing that can happen consistent with relativity and quantum mechanics is for it to become an excited nucleon with roughly the same energy and reduced longitudinal momentum and rapidity. It remains in that state inside the nucleus because the uncertainty principle and time dilation prevent it from fragmenting into particles until it is well outside the nucleus. This feature immediately eliminates the possibility of 
a cascade in the nucleus from the {rescattering} of the secondary products. 
If one makes the further assumptions that an excited nucleon interacts with the same 
cross section as an unexcited nucleon and that the successive collisions 
of the excited nucleon do not affect the excited state or its eventual 
fragmentation products~\cite{midFrankel}, this leads to the conclusion (c. 1977) that the elementary process for particle 
production in nuclear collisions is the excited nucleon, and to the prediction 
that the multiplicity in nuclear interactions should be proportional to 
the total number of projectile and target participants, rather than to the 
total number of collisions, or $R_A=N_{\rm part-pA}/N_{\rm part-pp}=(1+\nu)/2$, as observed.  This is called the Wounded Nucleon Model (WNM)~\cite{WNM} and, in the common usage, Wounded Nucleons (WN) are called participants.  

\begin{figure}[!h] 
      \centering
      \includegraphics[width=0.182\textwidth,angle=-1]{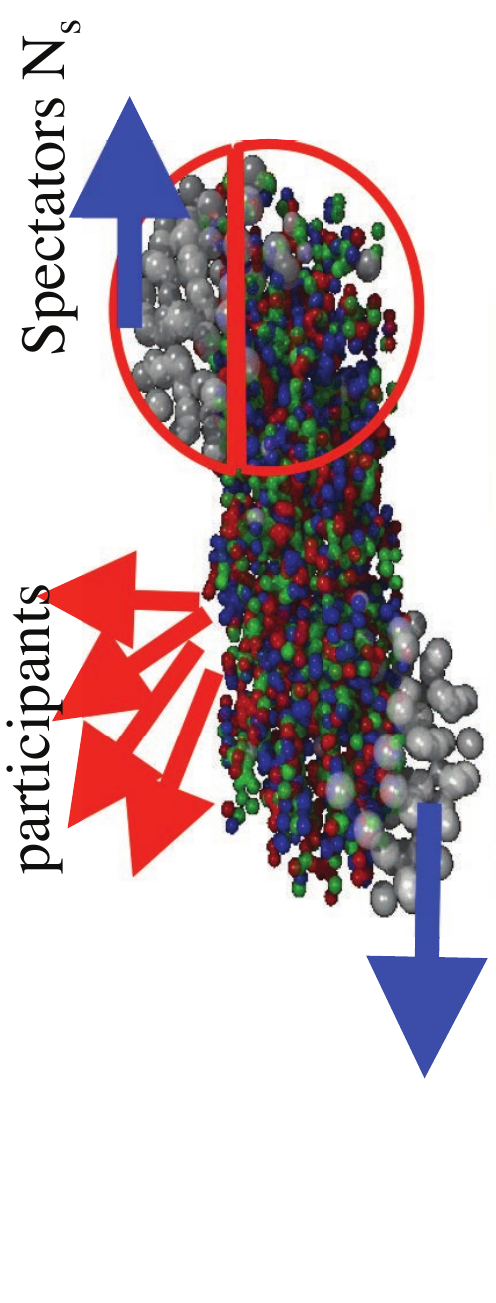}\hspace*{3pc}
      \includegraphics[width=0.406\textwidth,angle=-0.2]{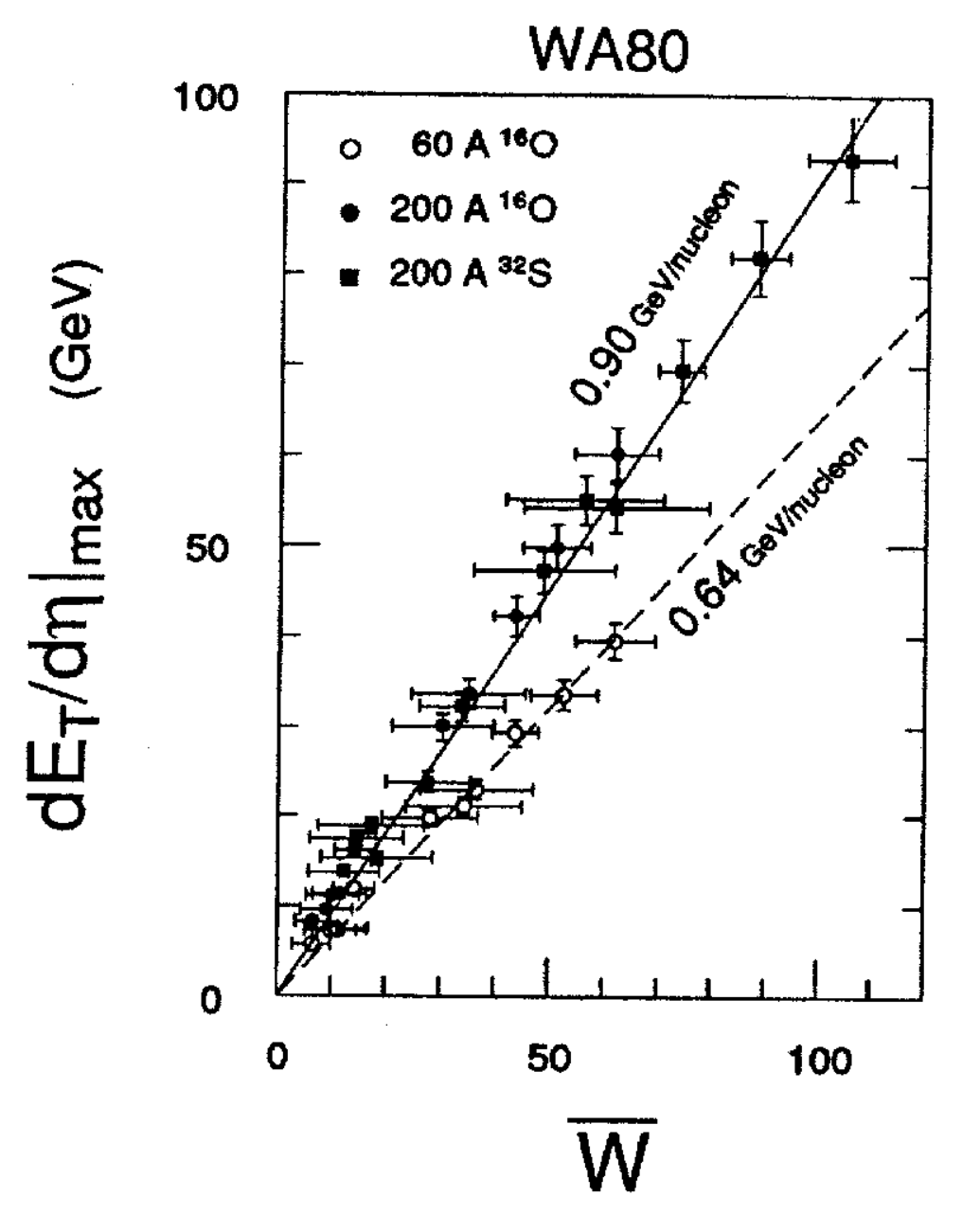} 
      \caption[]{a) (left) Schematic of an $A+A$ collision along the vertical axis with the spectators and participants indicated. b) (right) Measurement by WA80 in $O+A$ and $S+A$ collisions at CERN~\cite{WA80PRC44} of $d\Et/d\eta|_{\rm max}$ at mid-rapidity as a  function of $\mean{\Npart}\equiv {\overline W}$, where $A$ includes C, Al, Cu, Ag and Au.  }
      \label{fig:WA80WNM}
   \end{figure}

    In the fixed target experiments of the 1980's, the number of spectators (i.e. non-participants) $N_{\rm s}$  in a $B+A$ collion (Fig.~\ref{fig:WA80WNM}a) could be measured directly with a Zero Degree Calorimeter in the beam. This enabled an unambiguous measurement of the (projectile) participants, $N_{\rm part-proj}=B-N_{\rm s}$; and for a symmetric $A+A$ collision, the total number of participants, $\Npart=2 N_{\rm part-proj}=2 (A-N_{\rm s})$. In this way, the WA80 experiment~\cite{WA80PRC44} could directly verify the WNM, in beams of $^{16}O$ and $^{32}S$ of 60 and 200 $A\cdot$GeV/c on various nuclear targets, by the measured linearity of the mid-rapidity transverse energy density, $d\Et/d\eta|_{\rm max}$ with $\mean{\Npart}$ (Fig.~\ref{fig:WA80WNM}b).  
\section{Other Extreme Independent Models}
\begin{figure}[!h] 
      \centering
      \includegraphics[width=0.49\textwidth]{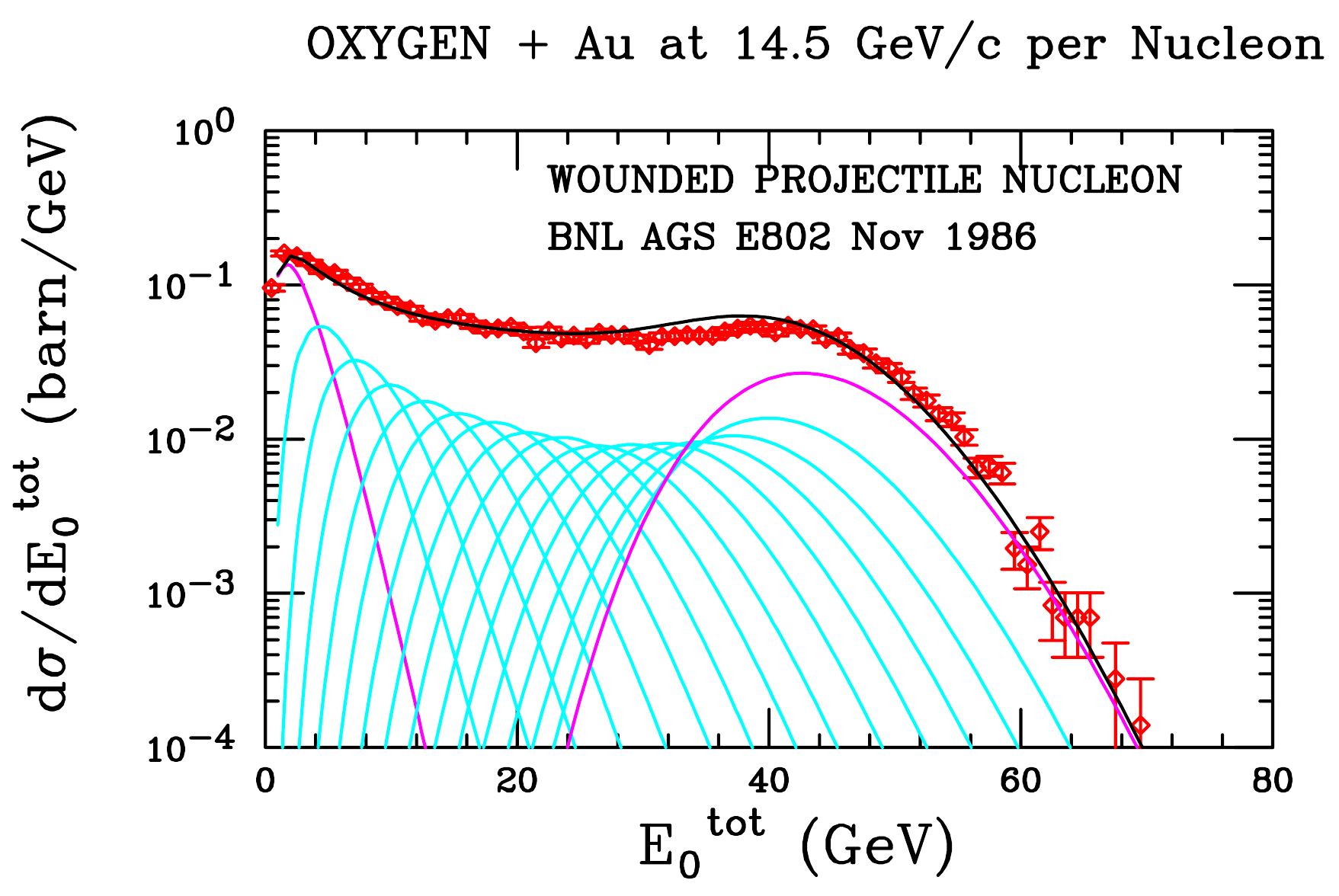}\hspace*{1pc} 
      \includegraphics[width=0.49\textwidth]{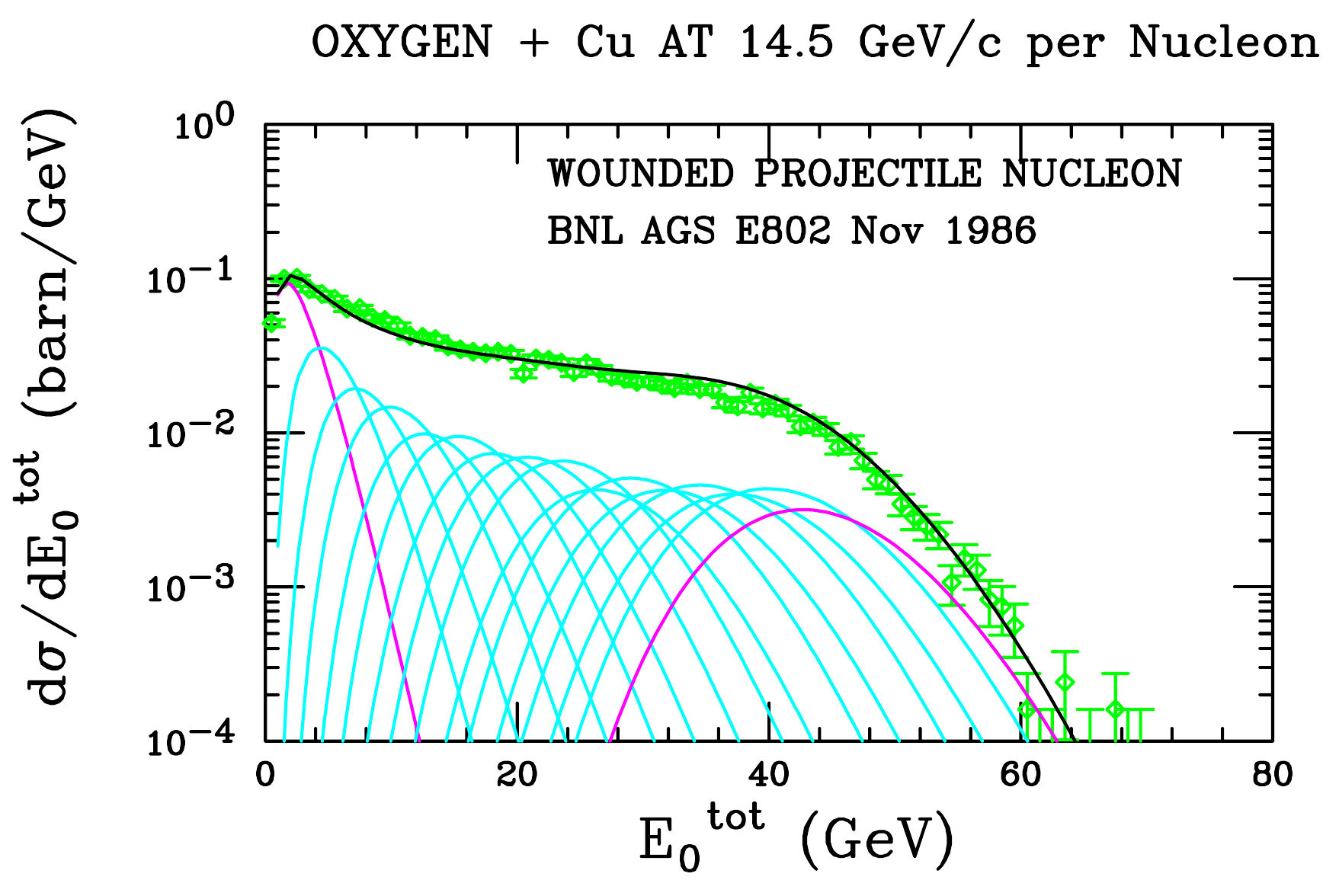}\vspace*{-1pc}
      \caption[]{Measured energy (E$_0^{\rm tot}$) distribution~\cite{E802ZPC38} in a Lead Glass Electromagnetic Calorimeter covering the full azimuth at mid-rapidity ($-0.47<\eta-y^{\rm cm}_{_{NN}}<0.72$) for 14.5 GeV/c $^{16}$O projectiles, where $y^{\rm cm}_{_{NN}}=1.72$. a) (left) O+Au, b) (right) O+Cu. Lines are from 1 to 16 convolutions of the measured p+Au distribution (see text).}
      \label{fig:famous}
   \end{figure}

Interestingly, at mid-rapidity, the WNM works well only at roughly \sqsn$\sim 20$ GeV where it was discovered. At lower \sqsn$\lsim$ 5.4 GeV, particle production is smaller than the WNM due to the large stopping~\cite{E866Akiba} with reduced transparency as shown by measurements by E802 at the BNL-AGS in an EM Calorimeter (Fig.~\ref{fig:famous})~\cite{E802ZPC38}. 
The large stopping at \sqsn$\lsim$ 5.4 GeV is indicated by the fact that the maximum energy in O+Cu is the same as in O+Au, even though the maximum thickness of a Cu nucleus is only $\approx 2/3$ that of Au. Also the upper edge of the O+Cu  spectrum is identical to that in O+Au but a factor of 6 lower in amplitude. Both these observations can be understood by representing the O+Cu and the O+Au spectra by sums of from 1 to 16 convolutions of the measured p+A spectrum according to the relative probability of the number of projectile participants, the Wounded Projectile Nucleon Model (WPNM)~\cite{FtPLB188,E802PLB197,E802ZPC38}. 

\begin{figure}[!h] 
      \centering
      \includegraphics[width=0.7\linewidth, height=0.75\linewidth]{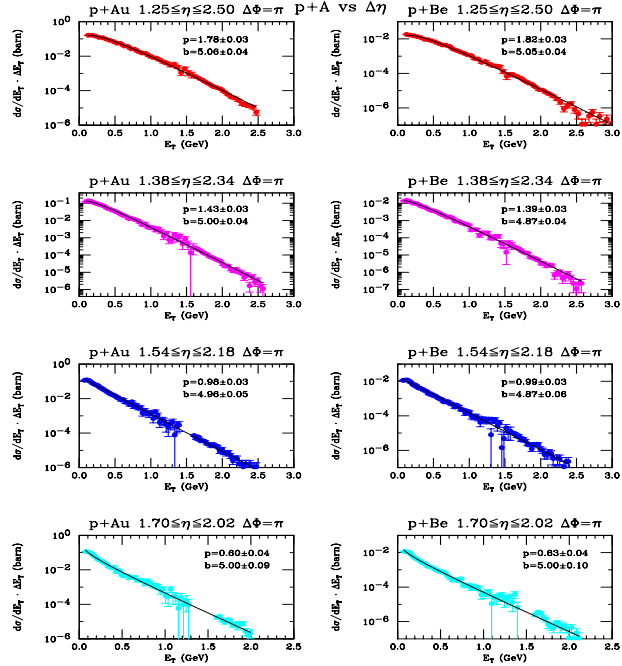}\vspace*{-2pc} 
      \caption[]{\Et distributions for p+Au (left) and p+Be (right) in a 14.6 GeV/c fixed target measurement~\cite{E802PRC63}. Adjacent plots have the same $\delta\eta$ interval which decreases from top to bottom. The solid lines are Gamma distribution fits with the parameters indicated. }\vspace*{-1pc}
      \label{fig:paubemid}
   \end{figure}

The stopping is more strikingly illustrated by the mid-rapidity \Et distributions for 14.6 GeV/c protons on Be and Au targets (Fig.~\ref{fig:paubemid})~\cite{E802PRC63}, 
which vary in shape as a function of the laboratory rapidity interval $\delta\eta$, but in each $\delta\eta$ interval, the p+Be and p+Au spectra are identical in shape. There is no evidence of multiple collisions in the target! This confirms a previous observation that the pion distribution from a second collision is shifted in rapidity by $\delta\eta>0.8$ which is out of the aperture of the \Et measurent. For the same reason, the Au+Au spectra~\cite{E802PRC63} are also well represented by the WPNM, a sum of convolutions of the measured p+Au (or p+Be) spectra. 
   
\begin{figure}[!h] 
      \centering
      \includegraphics[width=0.9\linewidth,angle=0.2]{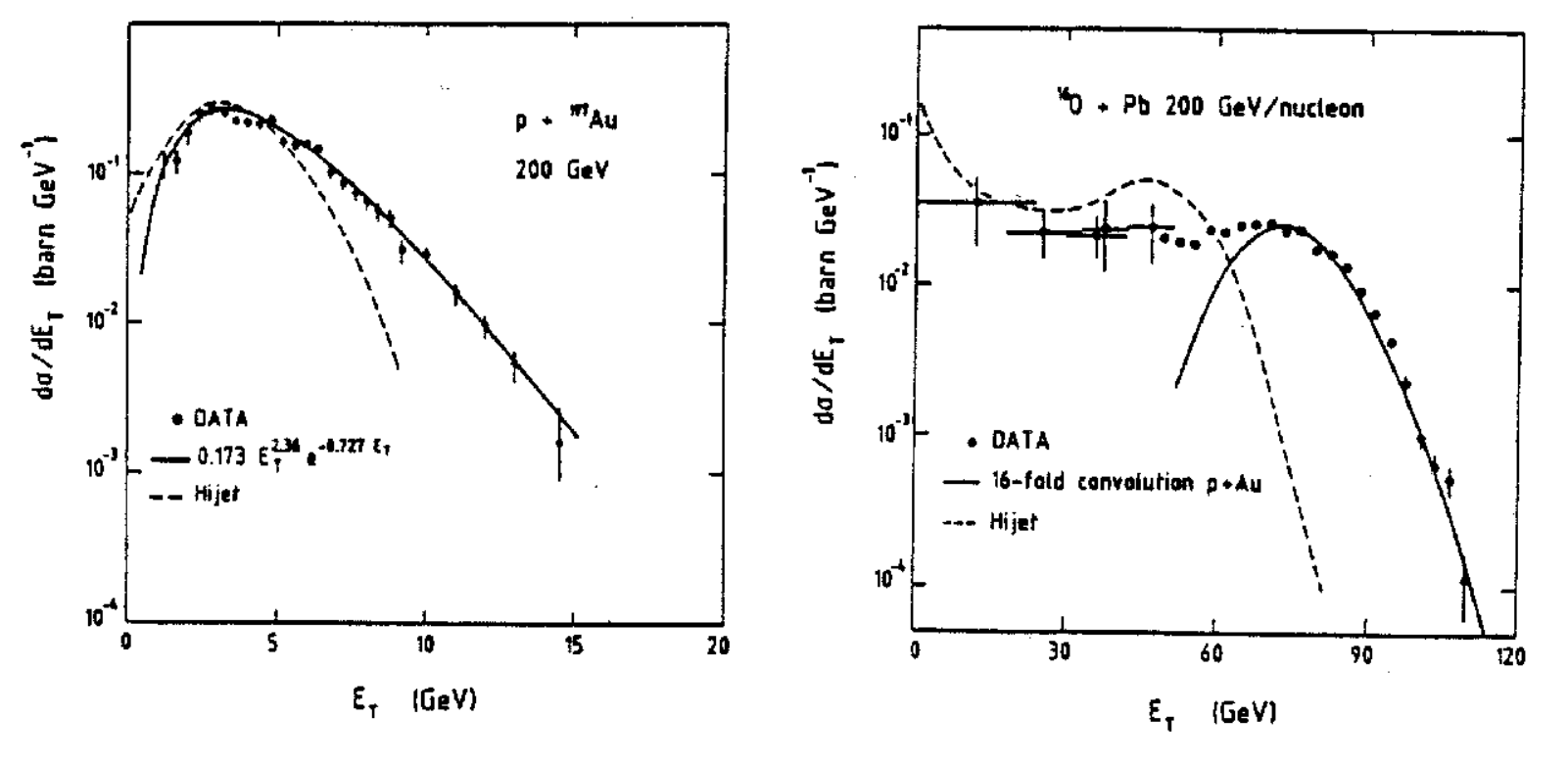} 
      \caption[]{Mid-rapidity \Et spectra of p+Au (left) and $^{16}$O+Pb from 200 GeV/c per nucleon fixed target measurements~\cite{NA35PLB184}. Solid lines are fit to the p+Au spectrum (left) and 16 fold convolution of the fit (right).  } 
      \label{fig:NA35pAOA}
   \end{figure}

It is interesting to note that the representation of the upper edge of the mid-rapidity O+Pb spectrum by 16 convolutions of the measured p+Au spectrum at $\sqsn\sim19.4$ GeV was first shown by NA35 at CERN (Fig.~\ref{fig:NA35pAOA})~\cite{NA35PLB184}, which inspired the WPNM~\cite{FtPLB188}, but does not contradict the WNM because a `centrality selected' p+Au spectrum was used~\cite{NA35ZPC38}, not the minimum bias p+Au spectrum. 

\begin{figure}[!h] 
      \centering
      \includegraphics[width=0.79\textwidth]{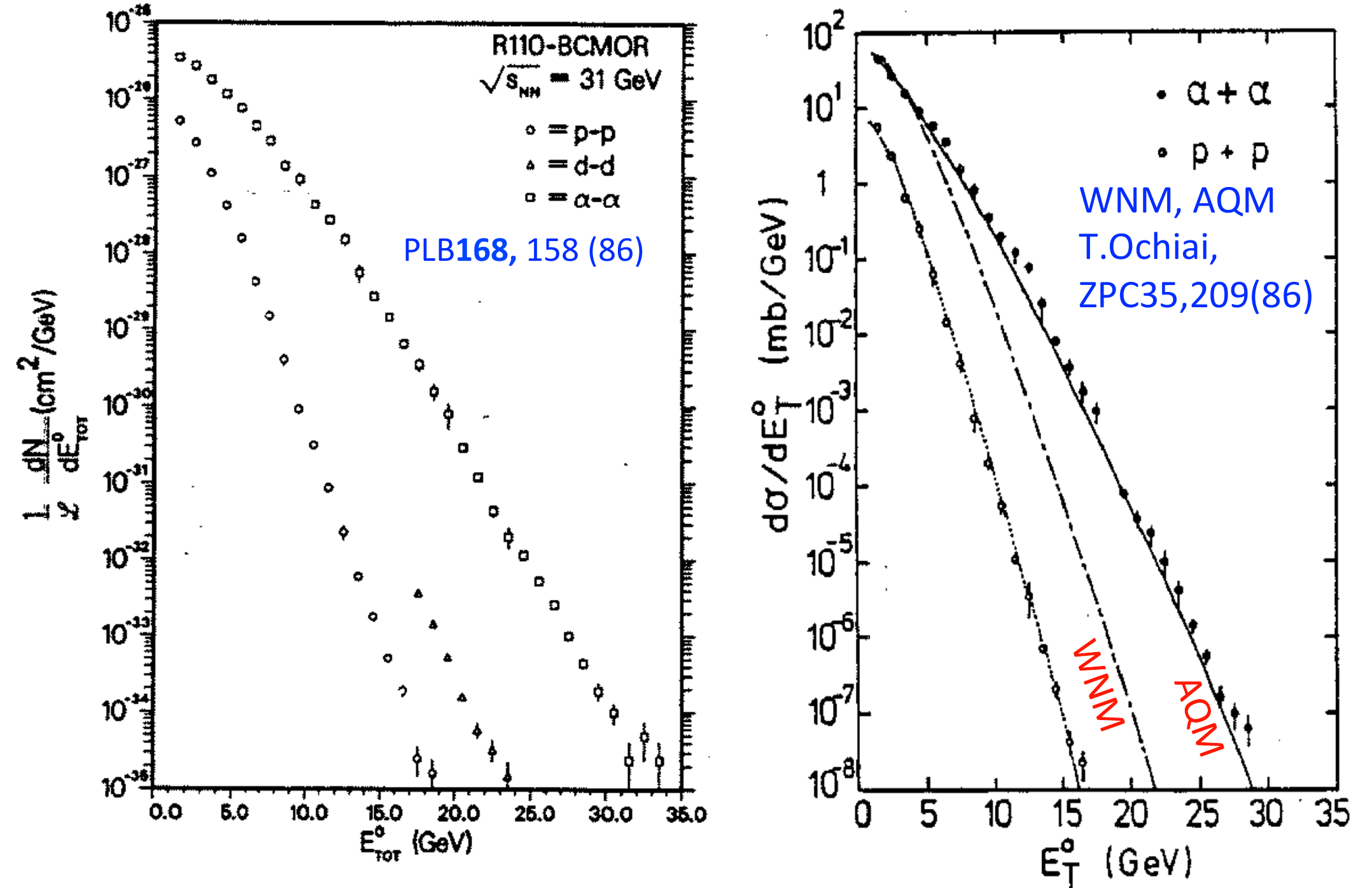} 
      \caption[]{a) (left) Total neutral energy $E^{0}_{\rm TOT}$  and b) (right) total neutral transverse energy $E^{0}_{\rm T}$  measured in an EM calorimeter at mid-rapidity $|\Delta y|<0.9$ , $\Delta \phi=1.8\pi$, in p+p, d+d, $\alpha+\alpha$ collisions at \sqsn=31 GeV at the CERN ISR~\cite{BCMOR-alfalfa}. Lines on the right are fits to the p-p data plus WNM and AQM calculations~\cite{OchiaiZPC35} as indicated.}  
      \label{fig:alfalfa}
   \end{figure}

For \sqsn$\geq 31$ GeV, particle production is larger than the WNM~\cite{BCMOR-alfalfa, AFSET89}; and a new model, the Additive Quark Model (AQM)~\cite{AQMPRD25,OchiaiZPC35}, which is equivalent to a wounded projectile quark (color-string) model, has been used successfully (Fig.~\ref{fig:alfalfa}). 
   
   All three of the above models as well as another to be described below are of the type referred to as ``Extreme Independent Models'' in which the effect of the nuclear geometry of the interaction can be calculated independently of the dynamics of particle production, which can be derived from experimental measurements, usually the p+p (or p+A) measurement in the same detector which typically follows a $\Gamma$ distribution.  A characteristic of these models can be seen in Fig.~\ref{fig:alfalfa}b in which the WNM line is parallel to the p+p spectrum because the 8-wounded-nucleon (4 collision) nuclear geometry is exhausted so that the shape of the 4-th convolution of the p+p spectrum (a $\Gamma$ distribution with the same exponential slope) dominates.~\footnote{The p+p fit is a $\Gamma$ distribution, $[b/\Gamma(p)](bE)^{p-1} e^{-bE}$,  with $p=2.50\pm 0.06$, $b=1.41\pm0.01$ GeV$^{-1}$. The 4th convolution is a $\Gamma$ distribution with $p\rightarrow 4p$ same $b$. }
\section{Measurements at RHIC and LHC-Ions}
\begin{figure}[!h] 
      \centering
      \includegraphics[width=0.99\textwidth]{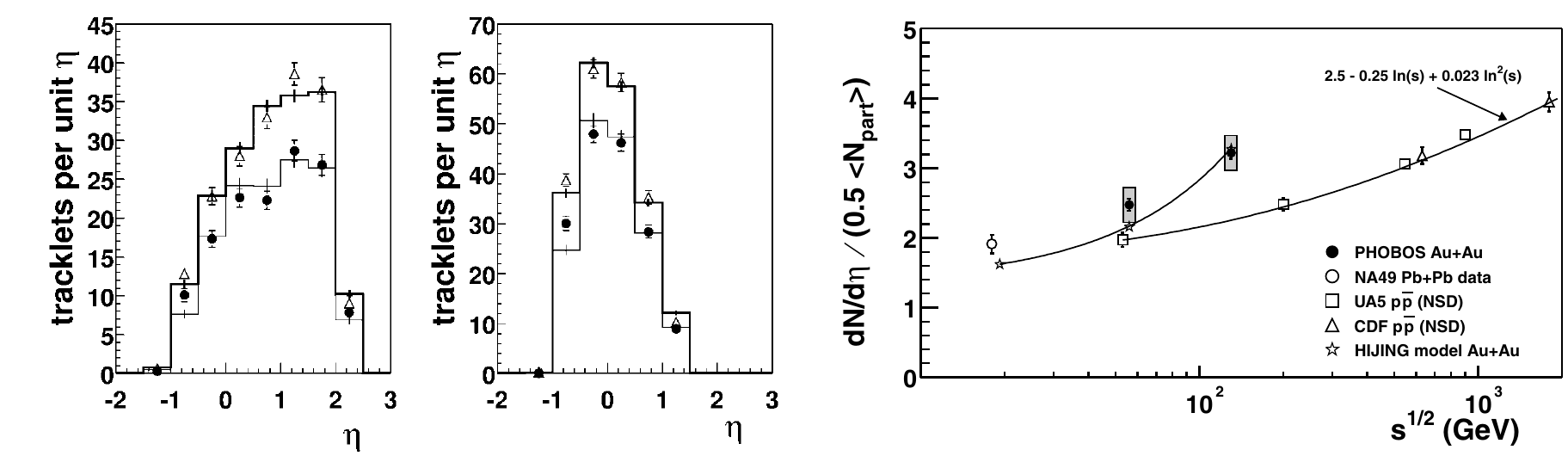} %%0.89
      \caption[]{a) (left) Pseudo-rapidity density of ``tracklets'' in ``SPEC'' and ``VTX'' elements of PHOBOS~\cite{PHOBOSPRL85} at $\sqrt{s_{NN}}=56$ (circles) and 130 GeV (triangles). b) (right) Measurements of $dN_{\rm ch}^{AA}/d\eta/(0.5 \mean{\Npart})$ compared to p+p values as a function of $\sqrt{s}$. }
      \label{fig:PhobosFirst}\vspace*{-1pc}
   \end{figure}

Wit's PHOBOS experiment produced the first published measurement at RHIC~\cite{PHOBOSPRL85}. The speciality of PHOBOS was its large rapidity coverage  (Fig.~\ref{fig:PhobosFirst}a); but a legacy of PHOBOS is that this first publication established the standard of quoting the mid-rapidity multiplicity distribution in A+A collisions as $dN_{\rm ch}^{AA}/d\eta/(0.5 \mean{\Npart})$ which would equal the value in p-p collisions at the same $\sqrt{s}$ if the data followed the WNM.  
The deviation from the WNM is shown clearly in Fig.~\ref{fig:PhobosFirst}b in which $dN_{\rm ch}^{AA}/d\eta/(0.5 \mean{\Npart})$ is above the p+p value at the same $\sqrt{s}$ for $\sqrt{s_{NN}}=56$ GeV and well above the p+p value for $\sqrt{s_{NN}}=130$ GeV.   

\begin{figure}[!b] 
      \centering
      \includegraphics[width=0.99\textwidth]{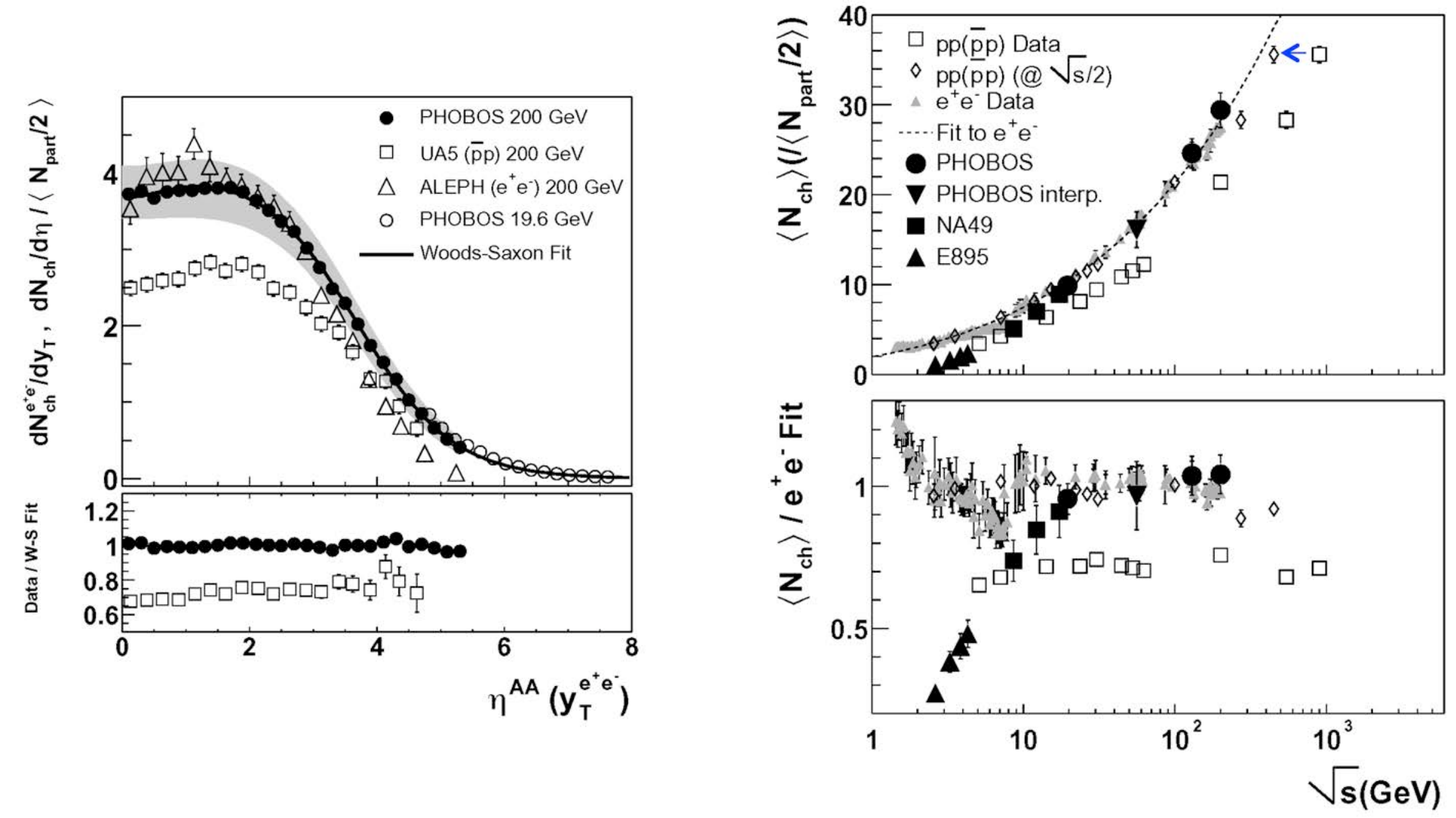} 
      \caption[]{a) (left) $dN_{\rm ch}^{AA}/d\eta/(0.5 \mean{\Npart})$~\cite{PHOBOSPRC74} for Au+Au collisions at $\sqrt{s_{NN}}=200$ GeV compared to $e^+ e^-$ and p+$\overline{\rm p}$ collisions at the same $\sqrt{s}$. b) (right) Total charge multiplicity in $e^+ e^-$, pp (p$\overline{\rm p}$), and $N_{\rm ch}^{AA}/d\eta/(0.5 \mean{\Npart})$ in A+A collisions as a function of $\sqrt{s}$. }
      \label{fig:Phobos-Nino}\vspace*{-1pc}
   \end{figure}

\begin{figure}[!b] 
      \centering
      \includegraphics[width=0.33\textwidth]{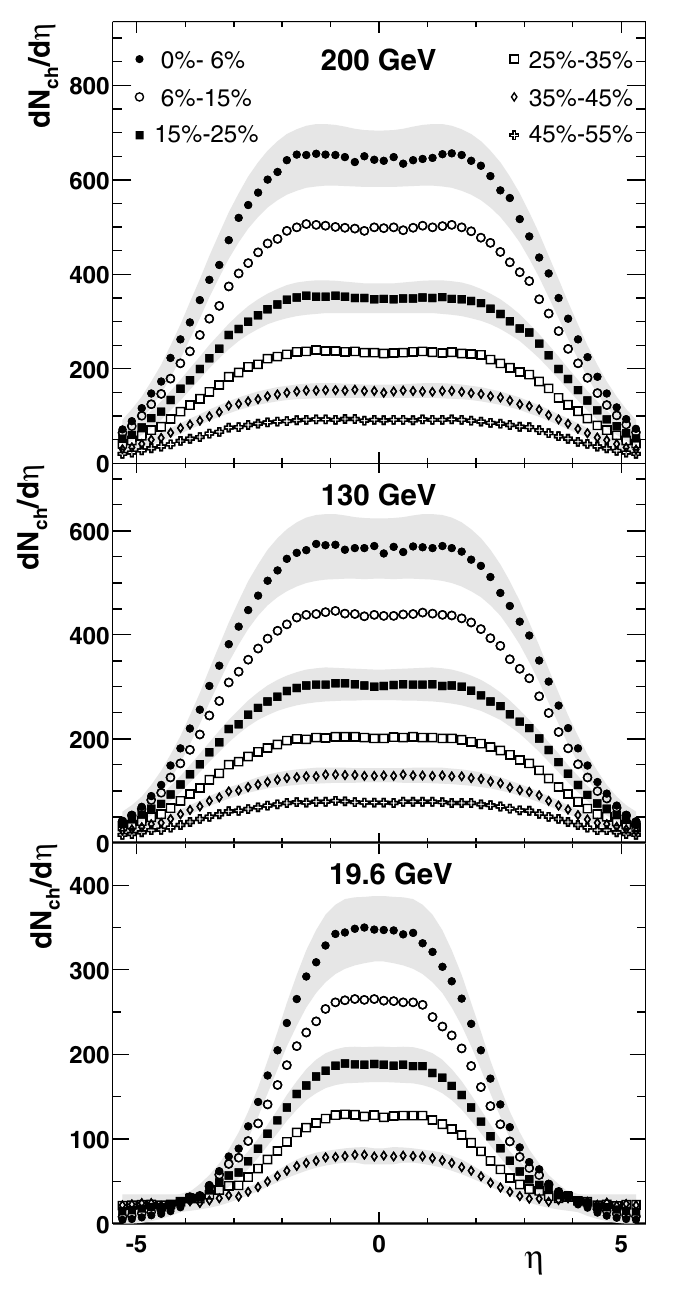}\hspace*{1pc} 
      \includegraphics[width=0.49\textwidth]{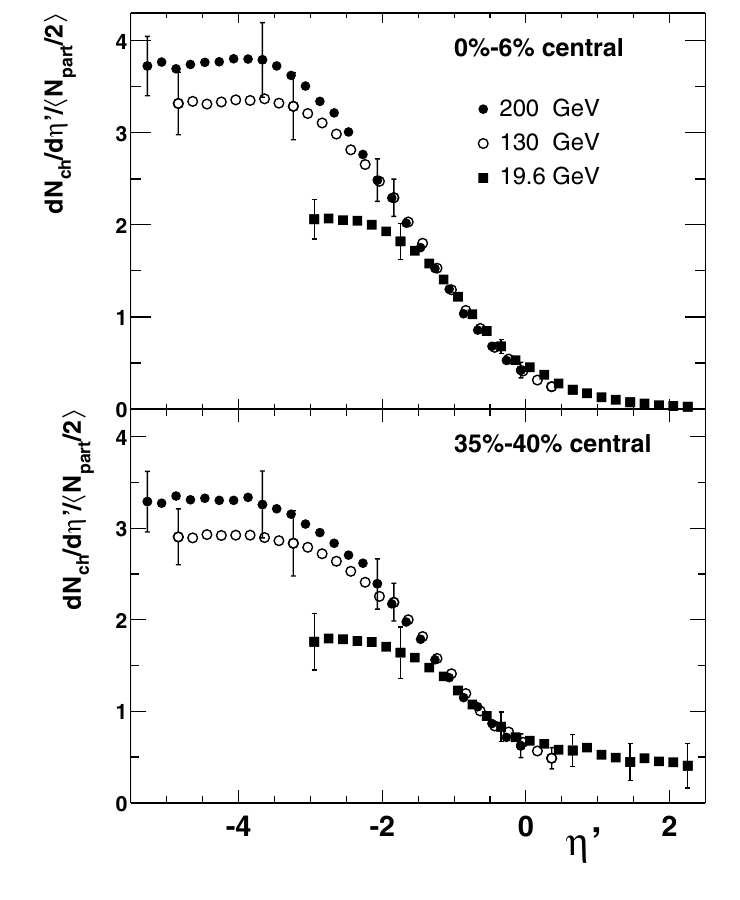}
      \caption[]{a) (left) The charged particle pseudorapidity distribution, $d\Nch/d\eta$ for Au+Au collisions at $\sqrt{s_{NN}}=19.6$, 130 and 200 GeV for specified centrality as indicated~\cite{PHOBOSPRL91}. b) $dN_{\rm ch}^{AA}/d\eta'/(0.5 \mean{\Npart})$ as a function of $\eta'=\eta-y_{\rm beam}$. }
      \label{fig:LimFrag}
   \end{figure}

	PHOBOS has also made some interesting observations regarding two well-established scaling laws in p+p physics, namely, the ``effective energy'', or ``leading particle effect''~\cite{BasilePLB95}, (Fig.~\ref{fig:Phobos-Nino})~\cite{PHOBOSPRC74}           
and ``limiting fragmentation''~\cite{BeneckePR188} (Fig.~\ref{fig:LimFrag})~\cite{PHOBOSPRL91}.

The ``leading particle'' effect is shown in Fig.~\ref{fig:Phobos-Nino}b: the total charged multiplicity in p+p collisions plotted at half the c.m. energy, i.e. at $\sqrt{s}/2$, is the same as that in $e^+ e^-$ collisions at $\sqrt{s}$. The reason is that the available or effective energy for making secondaries in p+p collisions is only 1/2 the c.m. energy because the two leading protons carry away half the energy~\cite{BasilePLB95}. PHOBOS has discovered~\cite{PHOBOSPRC74} that the leading particle effect vanishes in A+A collisions, i.e. $N_{\rm ch}^{AA}/d\eta/(0.5 \mean{\Npart})$ in A+A is the same as that in $e^+ e^-$ at the same value of $\sqrt{s}$ (Fig.~\ref{fig:Phobos-Nino}), because the leading protons in the first collision lose the rest of their energy via particle production on subsequent collisions. 

Figure~\ref{fig:LimFrag}~\cite{PHOBOSPRL91} shows that the width of the $d\Nch/d\eta$ distribution for Au+Au collisions narrows for more central collisions, with relatively fewer particles at beam and target rapidity; and that $dN_{\rm ch}^{AA}/d\eta'/(0.5 \mean{\Npart})$ measured with respect to the beam rapidity, $\eta'=\eta-y_{\rm beam}$, is universal in Au+Au collisions as it is in p+p collisions.  
\pagebreak

\begin{figure}[!t] 
      \centering
      \includegraphics[width=0.455\textwidth]{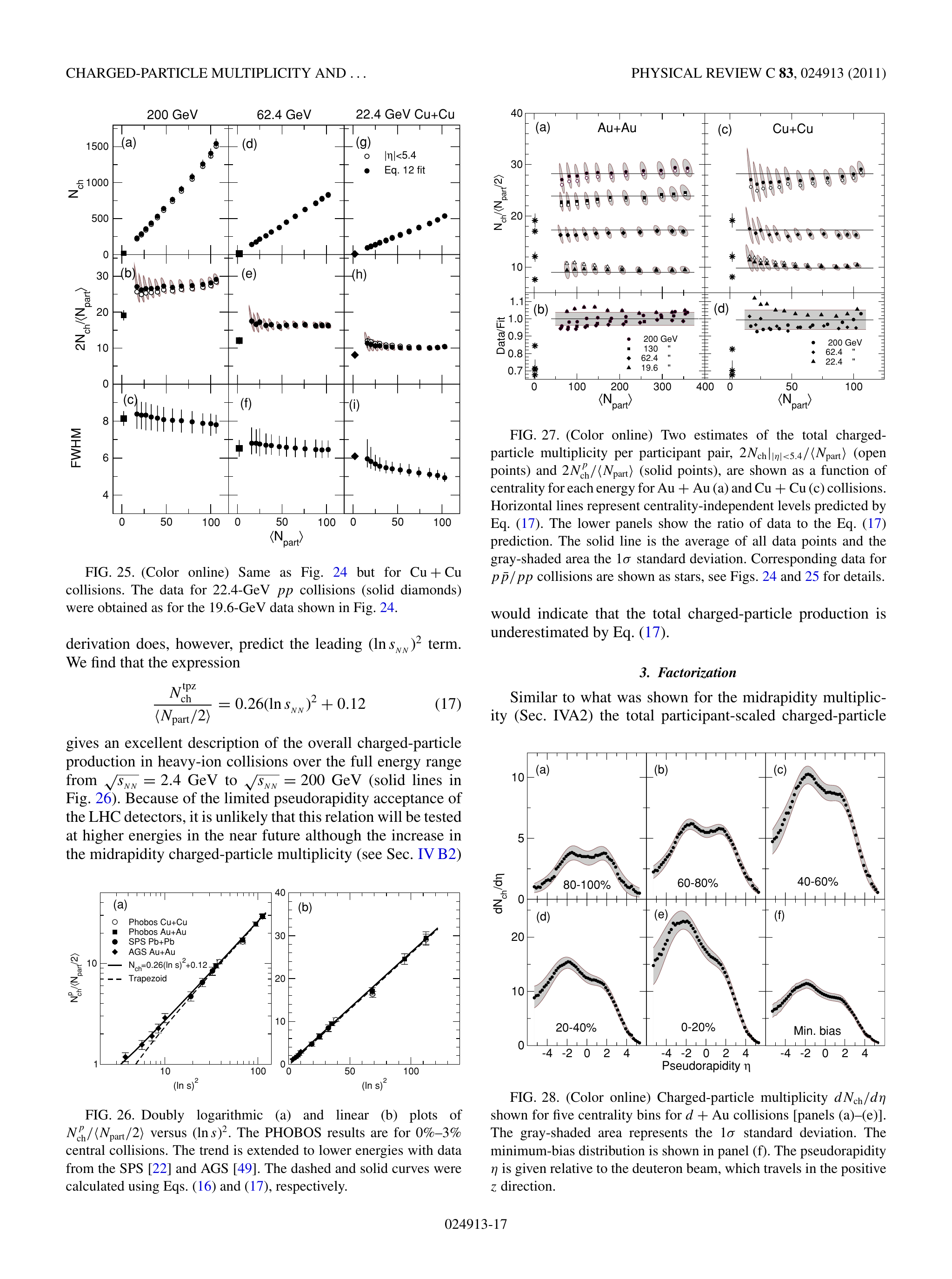}\hspace*{1pc} 
      \includegraphics[width=0.49\textwidth]{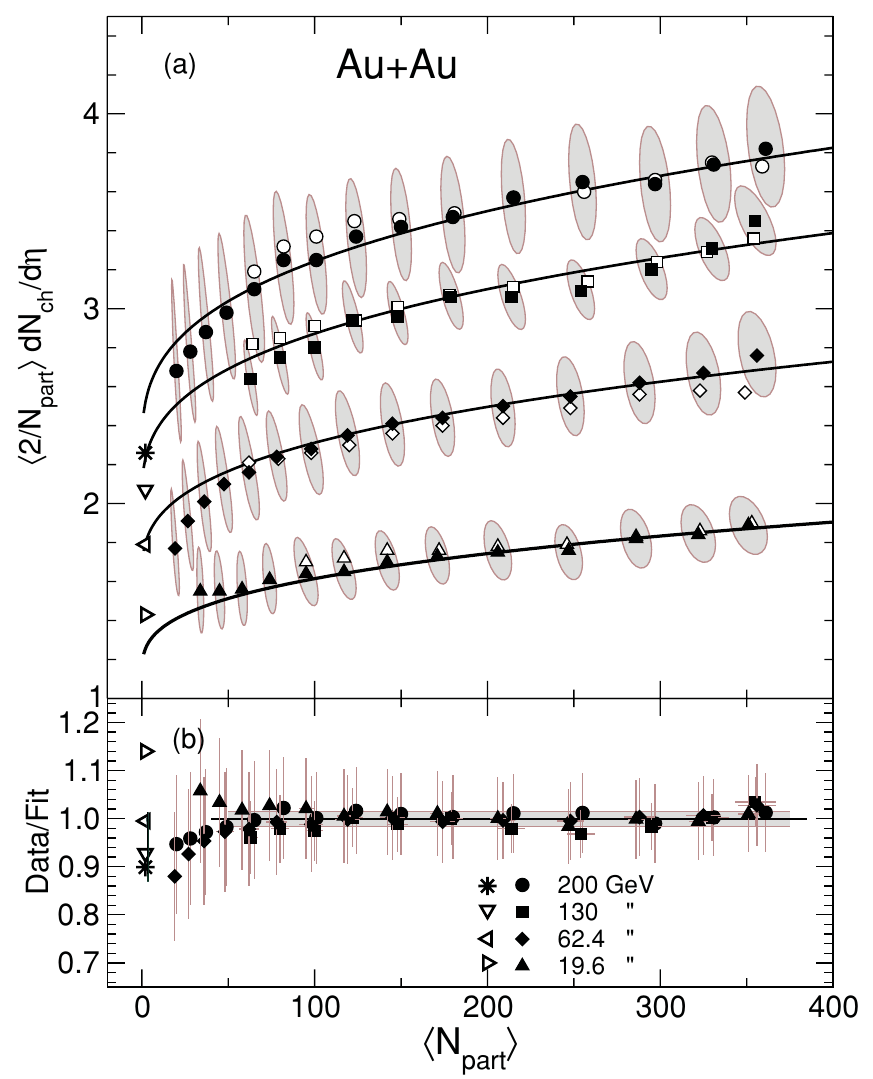}
      \caption[]{a) (left) Total charged multiplicity per participant pair $\Nch/(0.5 \mean{\Npart})$ (solid points) and  over the range $|\eta|<5.4$ (open points) as a function of centrality ($\mean{\Npart}$) for several values of $\sqrt{s_{NN}}$ indicated~\cite{PHOBOSPRC83}. b) (right) Midrapidity $dN_{\rm ch}^{AA}/d\eta/(0.5 \mean{\Npart})$ as a function of $\mean{\Npart}$. }
      \label{fig:Phobos-dndeta}
   \end{figure}

\begin{figure}[!h] 
      \centering
      \includegraphics[width=0.99\textwidth]{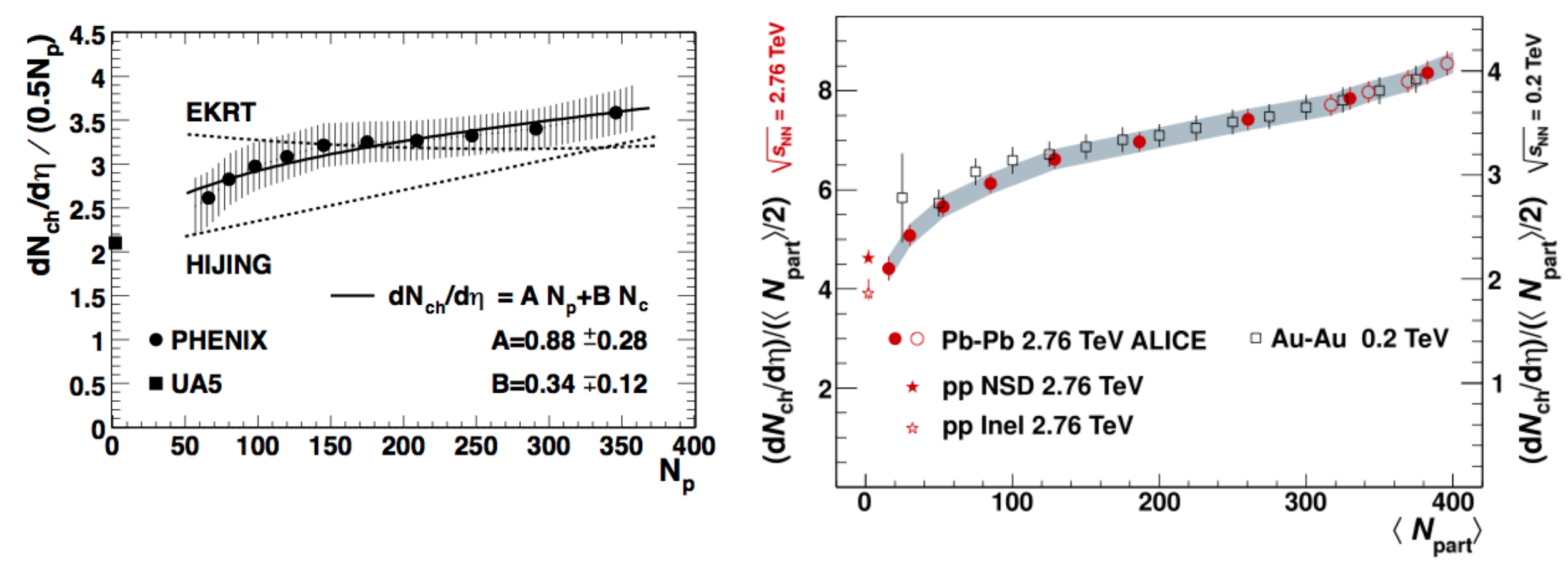} 
      \caption[]{a) (left) PHENIX~\cite{Adcox:2000sp} measurement of $dN_{\rm ch}^{AA}/d\eta/(0.5 \mean{\Npart})$ vs. $\mean{\Npart}$ at midrapidity in Au+Au collisions at $\sqrt{s_{NN}}=130$ GeV. b) (right) $dN_{\rm ch}^{AA}/d\eta/(0.5 \mean{\Npart})$ vs. $\mean{\Npart}$ at midrapidity, for Pb+Pb collisions at $\sqrt{s_{NN}}=2.76$ TeV and Au+Au collisions at $\sqrt{s_{NN}}=0.2$ TeV~\cite{ALICEPRL106}. The scale for the lower-energy data (right side) differs by a factor of 2.1 from the scale of the higher-energy data (left-side).}
      \label{fig:LHCvsRHIC}
   \end{figure}

In PHOBOS' final multiplicity paper~\cite{PHOBOSPRC83}, the systematics of charged particle production at RHIC were neatly summarized in two panels (Fig.~\ref{fig:Phobos-dndeta}).  
At RHIC, for \sqsn from  19.6 to 200 GeV (Fig.~\ref{fig:Phobos-dndeta}a), the WNM works in Au+Au collisions for the total multiplicity, $\Nch/(0.5 \mean{\Npart})$ and over the range $|\eta|<5.4$, while at mid-rapidity (Fig.~\ref{fig:Phobos-dndeta}b), the WNM fails---the multiplicity density per participant pair, $dN_{\rm ch}^{AA}/d\eta/(0.5 \mean{\Npart})$, increases with increasing number of participants, in a characteristic shape that was first observed by PHENIX~\cite{Adcox:2000sp} (Fig.~\ref{fig:LHCvsRHIC}a). 

With the recent startup of Heavy Ion Physics at the LHC, a huge increase in $\sqrt{s_{NN}}$ of more than an order of magnitude from 200 GeV to 2.76 TeV became avalable. The results for $dN_{\rm ch}^{AA}/d\eta/(0.5 \mean{\Npart})$ in Pb+Pb at midrapidity as first published by the ALICE experiment~\cite{ALICEPRL106} (Fig.~\ref{fig:LHCvsRHIC}b) were astounding: the ratio of  
$dN_{\rm ch}^{AA}/d\eta/(0.5 \mean{\Npart})$ from LHC to RHIC is simply a factor of 2.1 in every centrality bin. The LHC and RHIC measurements lie one on top of each other by simple scaling by a factor of 2.1. 

\begin{figure}[!h] 
      \centering
      \includegraphics[width=0.44\textwidth]{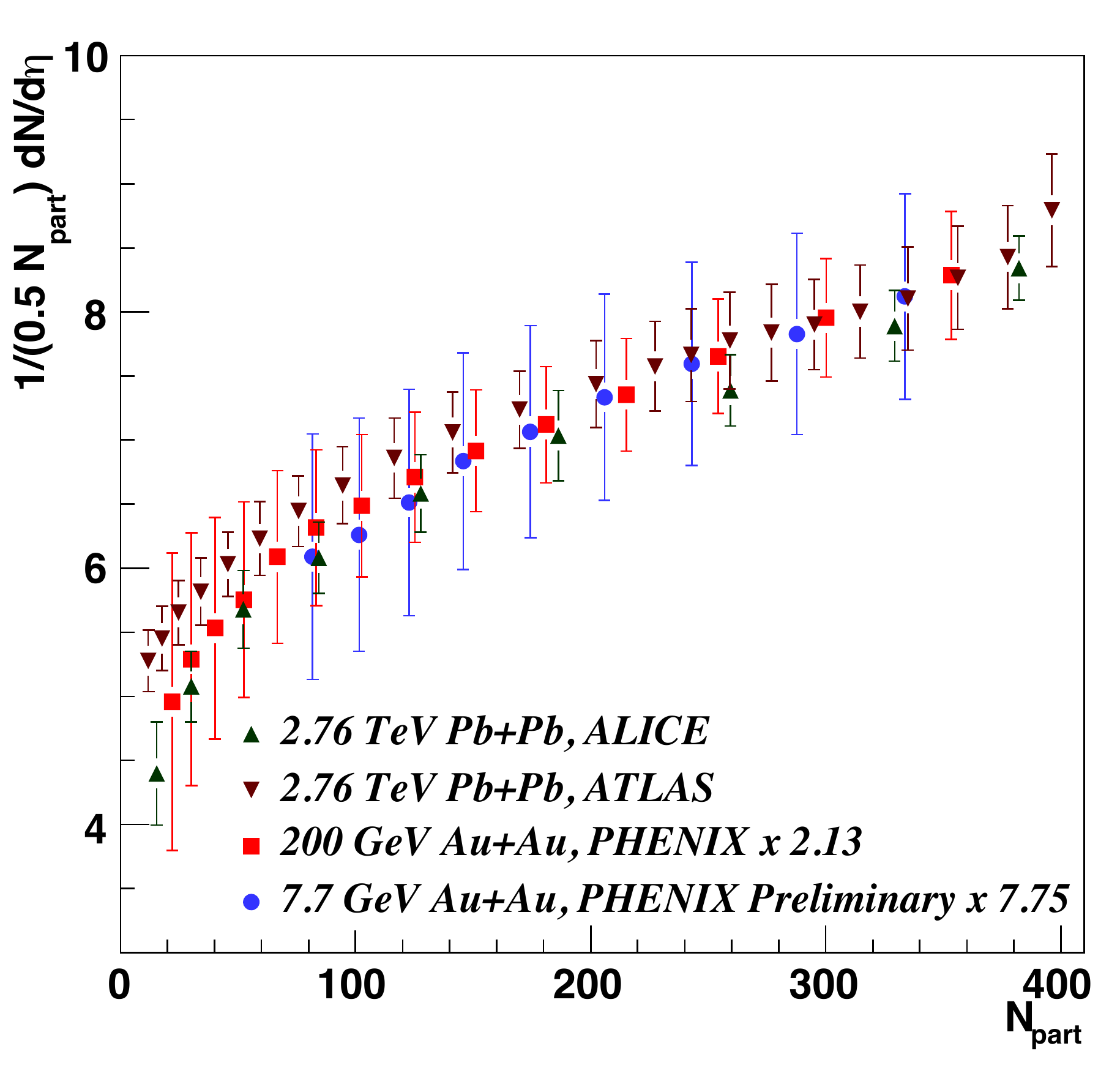}\hspace*{1pc} 
      \includegraphics[width=0.55\textwidth]{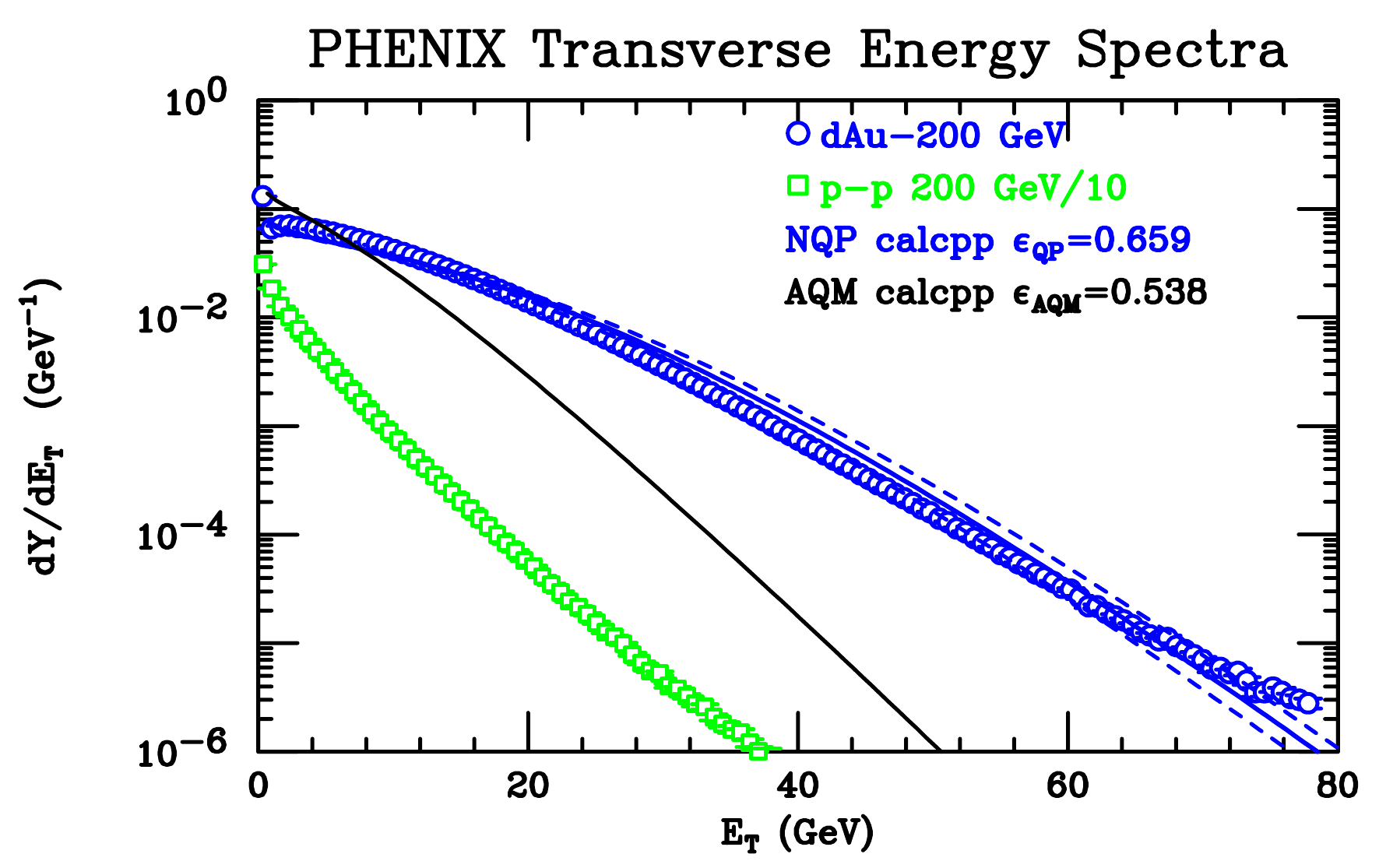}
      \caption[]{a) (left) $dN_{\rm ch}^{AA}/d\eta/(0.5 \mean{\Npart})$ data from Fig.~\ref{fig:LHCvsRHIC}b   
with new PHENIX preliminary measurement in Au+Au at \sqsn=0.0077 TeV. b) (right) PHENIX preliminary measurement of \Et distributions for p+p and d+Au at \sqsn=200 GeV together with calculations of the d+Au spectrum based on the AQM (color-strings) and the number of constituent-quark participants (NQP). }
      \label{fig:NQP}
   \end{figure}

In Fig.~\ref{fig:NQP}a new results from Au+Au collisions at \sqsn=7.7 GeV at RHIC scaled by a factor of 7.7 also lie on this curve. The identical shape of the centrality dependence of charged particle production for all \sqsn indicates that the dominant effect is the nuclear geometry of the A+A collision. 

It has been shown~\cite{EreminVoloshinPRC67,NouicerEPJC49} using PHOBOS Au+Au data from RHIC, and also discussed for other RHIC data~\cite{DeBhattPRC71}, that the geometry represents the number of constituent-quark participants, \Nqp : i.e. in Au+Au collisions, the mid-rapidity $\mean{dN_{\rm ch}^{AA}/d\eta}$ as a function centrality is linearly proportional to the number of constituent-quark participants (NQP). However, for symmetric systems, the NQP cannot be distinguished from the number of color-strings, the Additive Quark Model (AQM)~\cite{AQMPRD25,Bialas2008}. In the AQM, only one color string can be attached to a wounded quark. Thus, for asymmetric systems such as d+Au, the AQM is a ``wounded projectile quark" model since in this model, only 6 color strings can be attached between the wounded quarks in the d-projectile and Au-target, while the d+Au system can have many more quark participants. The PHENIX preliminary results (Fig.~\ref{fig:NQP}b) show that in fact it is the NQP not the AQM (color string model) that works. 

\section{Some Personal Observations}
I have spent a lot of my research time on measurements and ``Extreme Independent Models'' based on and inspired by Wit's work. I have also greatly enjoyed collaborating on E802 at the AGS with the MIT group: Lee Grodzins, Steve Steadman, George Stephans and many great MIT graduate students who are now among the leaders of the field. I am also impressed with the excellent group that Wit has built up here at MIT and the outstanding contributions that they have made to RHI physics both at RHIC on PHOBOS and now at LHC on CMS. However, I also must confess that my personal favorite RHIC result from the MIT group does not relate to \Npart or \Ncoll, but rather to ``triangular flow''~\cite{AlverRolandPRC81}. 

For the first 10 years of RHIC running and dating back to the Bevalac, all of the experts, with possibly one exception~\cite{TakahashiPRL103}, thought that the odd harmonics of collective flow vanished at mid-rapidity due to the symmetry of the source for $\phi\rightarrow \pi +\phi$, even though `head and shoulders' or ``Mach Cones'' were observed at $\Delta\phi=\pi\pm \pi/3$ which prevented me from understanding the two-particle correlations from di-jets~\cite{MJTCFRNC06}. In 2006 at the Correlations and Fluctuations conference in Florence, Italy, I was chair of a session at which Burak Alver presented some ideas of how to measure the fluctuations of $v_2$, $\sigma^2_{v_2}=\mean{v_2^2}-\mean{v_2}^2$, in PHOBOS~\cite{AlverCFRNC06}. Frankly I was skeptical about the method he proposed, and George Stephans was skeptical about my skepticism. However, in 2010~\cite{AlverRolandPRC81}, Burak and his Professor, Gunther Roland, realized that due to fluctuations in the collision geometry on an event by event basis, the eccentricity of participants on any given event did not respect the average symmetry, resulting in ``Triangular Flow'', or $v_3$, another legacy of MIT to RHI physics.

%%%MJT 022113-after discussion with SashaB, felt it necessary to emphasize that these models came with the first CERN and BNL fixed target runs



%% References


\bibliographystyle{model1a-num-names}
\bibliography{MJT-Busza}

\begin{thebibliography}{39}
\expandafter\ifx\csname natexlab\endcsname\relax\def\natexlab#1{#1}\fi
\providecommand{\url}[1]{\texttt{#1}}
\providecommand{\href}[2]{#2}
\providecommand{\path}[1]{#1}
\providecommand{\DOIprefix}{doi:}
\providecommand{\ArXivprefix}{arXiv:}
\providecommand{\URLprefix}{URL: }
\providecommand{\Pubmedprefix}{pmid:}
\providecommand{\doi}[1]{\href{http://dx.doi.org/#1}{\path{#1}}}
\providecommand{\Pubmed}[1]{\href{pmid:#1}{\path{#1}}}
\providecommand{\bibinfo}[2]{#2}
\ifx\xfnm\relax \def\xfnm[#1]{\unskip,\space#1}\fi
%Type = Article
\bibitem[{Busza et~al.(1975)}]{BuszaPRL34}
\bibinfo{author}{W.~Busza}, et~al., \bibinfo{journal}{Phys.~Rev.~Lett.}
  \bibinfo{volume}{34} (\bibinfo{year}{1975}) \bibinfo{pages}{836--839}.
  \URLprefix \url{http://link.aps.org/doi/10.1103/PhysRevLett.34.836}.
%Type = Article
\bibitem[{Elias et~al.(1980)}]{EliasPRD22}
\bibinfo{author}{J.~E. Elias}, et~al., \bibinfo{journal}{Phys.~Rev.~D}
  \bibinfo{volume}{22} (\bibinfo{year}{1980}) \bibinfo{pages}{13--35}.
  \URLprefix \url{http://link.aps.org/doi/10.1103/PhysRevD.22.13}.
%Type = Article
\bibitem[{Halliwell et~al.(1977)}]{HalliwellPRL39}
\bibinfo{author}{C.~Halliwell}, et~al., \bibinfo{journal}{Phys.~Rev.~Lett.}
  \bibinfo{volume}{39} (\bibinfo{year}{1977}) \bibinfo{pages}{1499--1502}.
  \URLprefix \url{http://link.aps.org/doi/10.1103/PhysRevLett.39.1499}.
%Type = Article
\bibitem[{Fishbane and Trefil(1974{\natexlab{a}})}]{FishbaneTrefilPRD9}
\bibinfo{author}{P.~M. Fishbane}, \bibinfo{author}{J.~S. Trefil},
  \bibinfo{journal}{Phys.~Rev.~D} \bibinfo{volume}{9}
  (\bibinfo{year}{1974}{\natexlab{a}}) \bibinfo{pages}{168}. \URLprefix
  \url{http://link.aps.org/doi/10.1103/PhysRevD.9.168}.
%Type = Article
\bibitem[{Fishbane and Trefil(1974{\natexlab{b}})}]{FishbaneTrefilPLB51}
\bibinfo{author}{P.~M. Fishbane}, \bibinfo{author}{J.~S. Trefil},
  \bibinfo{journal}{Phys.~Lett.~B} \bibinfo{volume}{51}
  (\bibinfo{year}{1974}{\natexlab{b}}) \bibinfo{pages}{139}. \URLprefix
  \url{http://dx.doi.org/10.1016/0370-2693(74)90199-3}.
%Type = Article
\bibitem[{Gottfried(1974)}]{GottfriedPRL32}
\bibinfo{author}{K.~Gottfried}, \bibinfo{journal}{Phys.~Rev.~Lett.}
  \bibinfo{volume}{32} (\bibinfo{year}{1974}) \bibinfo{pages}{957}. \URLprefix
  \url{http://link.aps.org/doi/10.1103/PhysRevLett.32.957}.
%Type = Article
\bibitem[{Goldhaber(1973)}]{ASGoldhPRD7}
\bibinfo{author}{A.~S. Goldhaber}, \bibinfo{journal}{Phys.~Rev.~D}
  \bibinfo{volume}{7} (\bibinfo{year}{1973}) \bibinfo{pages}{765}. \URLprefix
  \url{http://link.aps.org/doi/10.1103/PhysRevD.7.765}.
%Type = Article
\bibitem[{Bia{\l}as and Czy{\.z}(1974)}]{BialasCzyzPLB51}
\bibinfo{author}{A.~Bia{\l}as}, \bibinfo{author}{W.~Czy{\.z}},
  \bibinfo{journal}{Phys.~Lett.~B} \bibinfo{volume}{51} (\bibinfo{year}{1974})
  \bibinfo{pages}{179}. \URLprefix
  \url{http://dx.doi.org/10.1016/0370-2693(74)90210-X}.
%Type = Article
\bibitem[{Andersson and Otterlund(1975)}]{BoIngvarNPB88}
\bibinfo{author}{B.~Andersson}, \bibinfo{author}{I.~Otterlund},
  \bibinfo{journal}{Nucl.~Phys.~B} \bibinfo{volume}{88} (\bibinfo{year}{1975})
  \bibinfo{pages}{349}. \URLprefix
  \url{http://dx.doi.org/10.1016/0550-3213(75)90286-2}.
%Type = Article
\bibitem[{Frankel(1993)}]{midFrankel}
\bibinfo{author}{S.~Frankel}, \bibinfo{journal}{Phys.~Rev.~C}
  \bibinfo{volume}{48} (\bibinfo{year}{1993}) \bibinfo{pages}{R2170}.
  \URLprefix \url{http://link.aps.org/doi/10.1103/PhysRevC.48.R2170}.
%Type = Article
\bibitem[{Bia{\l}as et~al.(1976)Bia{\l}as, B{l}eszy\'{n}ski, and
  Czy{\.z}}]{WNM}
\bibinfo{author}{A.~Bia{\l}as}, \bibinfo{author}{M.~B{l}eszy\'{n}ski},
  \bibinfo{author}{W.~Czy{\.z}}, \bibinfo{journal}{Nucl.~Phys.~B}
  \bibinfo{volume}{111} (\bibinfo{year}{1976}) \bibinfo{pages}{461--476}.
  \URLprefix \url{http://dx.doi.org/10.1016/0550-3213(76)90329-1}.
%Type = Article
\bibitem[{Albrecht et~al.(1991)}]{WA80PRC44}
\bibinfo{author}{R.~Albrecht}, et~al., \bibinfo{journal}{Phys.~Rev.~C}
  \bibinfo{volume}{44} (\bibinfo{year}{1991}) \bibinfo{pages}{2736--2752}.
  \URLprefix \url{http://link.aps.org/doi/10.1103/PhysRevC.44.2736}.
%Type = Article
\bibitem[{Ahle et~al.(1998)}]{E866Akiba}
\bibinfo{author}{L.~Ahle}, et~al., \bibinfo{journal}{Phys.~Rev.~C}
  \bibinfo{volume}{57} (\bibinfo{year}{1998}) \bibinfo{pages}{R466}. \URLprefix
  \url{http://link.aps.org/doi/10.1103/PhysRevC.57.R466}.
%Type = Article
\bibitem[{Remsberg et~al.(1988)}]{E802ZPC38}
\bibinfo{author}{L.~P. Remsberg}, et~al., \bibinfo{journal}{Z.~Phys.~C}
  \bibinfo{volume}{38} (\bibinfo{year}{1988}) \bibinfo{pages}{35--43}.
  \URLprefix \url{http://dx.doi.org/10.1007/BF01574512}.
%Type = Article
\bibitem[{Ftacnik et~al.(1987)Ftacnik, Kajantie, Pisutova, and
  Pisut}]{FtPLB188}
\bibinfo{author}{J.~Ftacnik}, \bibinfo{author}{K.~Kajantie},
  \bibinfo{author}{N.~Pisutova}, \bibinfo{author}{J.~Pisut},
  \bibinfo{journal}{Phys.~Lett.~B} \bibinfo{volume}{188} (\bibinfo{year}{1987})
  \bibinfo{pages}{279--282}. \URLprefix
  \url{http://dx.doi.org/10.1016/0370-2693(87)90021-9}.
%Type = Article
\bibitem[{Abbott et~al.(1987)}]{E802PLB197}
\bibinfo{author}{T.~Abbott}, et~al., \bibinfo{journal}{Phys.~Lett.~B}
  \bibinfo{volume}{197} (\bibinfo{year}{1987}) \bibinfo{pages}{285--290}.
  \URLprefix \url{http://dx.doi.org/10.1016/0370-2693(87)90385-6}.
%Type = Article
\bibitem[{Abbott et~al.(2001)}]{E802PRC63}
\bibinfo{author}{T.~Abbott}, et~al., \bibinfo{journal}{Phys.~Rev.~C}
  \bibinfo{volume}{63} (\bibinfo{year}{2001}) \bibinfo{pages}{064602}.
  \URLprefix \url{http://link.aps.org/doi/10.1103/PhysRevC.63.064602}.
%Type = Article
\bibitem[{Bamberger et~al.(1987)}]{NA35PLB184}
\bibinfo{author}{A.~Bamberger}, et~al., \bibinfo{journal}{Phys.~Lett.~B}
  \bibinfo{volume}{184} (\bibinfo{year}{1987}) \bibinfo{pages}{271}. \URLprefix
  \url{http://dx.doi.org/10.1016/0370-2693(87)90581-8}.
%Type = Article
\bibitem[{Heck et~al.(1988)}]{NA35ZPC38}
\bibinfo{author}{W.~Heck}, et~al., \bibinfo{journal}{Z.~Phys.~C}
  \bibinfo{volume}{38} (\bibinfo{year}{1988}) \bibinfo{pages}{19--34}.
  \URLprefix \url{http://dx.doi.org/10.1007/BF01574511}.
%Type = Article
\bibitem[{Angelis et~al.(1984)}]{BCMOR-alfalfa}
\bibinfo{author}{A.~L.~S. Angelis}, et~al., \bibinfo{journal}{Phys.~Lett.~B}
  \bibinfo{volume}{141} (\bibinfo{year}{1984}) \bibinfo{pages}{140}. \URLprefix
  \url{http://dx.doi.org/10.1016/0370-2693(84)90577-X}.
%Type = Article
\bibitem[{{\AA}kesson et~al.(1989)}]{AFSET89}
\bibinfo{author}{T.~{\AA}kesson}, et~al., \bibinfo{journal}{Phys.~Lett.~B}
  \bibinfo{volume}{231} (\bibinfo{year}{1989}) \bibinfo{pages}{359--364}.
  \URLprefix \url{http://dx.doi.org/10.1016/0370-2693(89)90676-X}.
%Type = Article
\bibitem[{Bialas et~al.(1982)Bialas, Czyz, and Lesniak}]{AQMPRD25}
\bibinfo{author}{A.~Bialas}, \bibinfo{author}{W.~Czyz},
  \bibinfo{author}{L.~Lesniak}, \bibinfo{journal}{Phys.~Rev.~D}
  \bibinfo{volume}{25} (\bibinfo{year}{1982}) \bibinfo{pages}{2328--2340}.
  \URLprefix \url{http://link.aps.org/doi/10.1103/PhysRevD.25.2328}.
%Type = Article
\bibitem[{Ochiai(1987)}]{OchiaiZPC35}
\bibinfo{author}{T.~Ochiai}, \bibinfo{journal}{Z.~Phys.~C} \bibinfo{volume}{35}
  (\bibinfo{year}{1987}) \bibinfo{pages}{209--214}. \URLprefix
  \url{http://dx.doi.org/10.1007/BF01408449}.
%Type = Article
\bibitem[{Back et~al.(2000)}]{PHOBOSPRL85}
\bibinfo{author}{B.~B. Back}, et~al., \bibinfo{journal}{Phys.~Rev.~Lett.}
  \bibinfo{volume}{85} (\bibinfo{year}{2000}) \bibinfo{pages}{3100--3104}.
  \URLprefix \url{http://link.aps.org/doi/10.1103/PhysRevLett.85.3100}.
%Type = Article
\bibitem[{Basile et~al.(1980)}]{BasilePLB95}
\bibinfo{author}{M.~Basile}, et~al., \bibinfo{journal}{Phys.~Lett.~B}
  \bibinfo{volume}{95} (\bibinfo{year}{1980}) \bibinfo{pages}{311--312}.
  \URLprefix \url{http://dx.doi.org/10.1016/0370-2693(80)90493-1}.
%Type = Article
\bibitem[{Back et~al.(2006)}]{PHOBOSPRC74}
\bibinfo{author}{B.~B. Back}, et~al., \bibinfo{journal}{Phys.~Rev.~C}
  \bibinfo{volume}{74} (\bibinfo{year}{2006}) \bibinfo{pages}{021902(R)}.
  \URLprefix \url{http://dx.doi.org/10.1103/PhysRevC.74.021902}.
%Type = Article
\bibitem[{Benecke et~al.(1969)Benecke, Chou, Yang, and Yen}]{BeneckePR188}
\bibinfo{author}{J.~Benecke}, \bibinfo{author}{T.~T. Chou},
  \bibinfo{author}{C.~N. Yang}, \bibinfo{author}{E.~Yen},
  \bibinfo{journal}{Phys.~Rev.} \bibinfo{volume}{188} (\bibinfo{year}{1969})
  \bibinfo{pages}{2159--2169}. \URLprefix
  \url{http://link.aps.org/doi/10.1103/PhysRev.188.2159}.
%Type = Article
\bibitem[{Back et~al.(2003)}]{PHOBOSPRL91}
\bibinfo{author}{B.~B. Back}, et~al., \bibinfo{journal}{Phys.~Rev.~Lett.}
  \bibinfo{volume}{91} (\bibinfo{year}{2003}) \bibinfo{pages}{052303}.
  \URLprefix \url{http://dx.doi.org/10.1103/PhysRevLett.91.052303}.
%Type = Article
\bibitem[{Alver et~al.(2011)}]{PHOBOSPRC83}
\bibinfo{author}{B.~Alver}, et~al., \bibinfo{journal}{Phys.~Rev.~C}
  \bibinfo{volume}{83} (\bibinfo{year}{2011}) \bibinfo{pages}{024913}.
  \URLprefix \url{http://dx.doi.org/10.1103/PhysRevC.83.024913}.
%Type = Article
\bibitem[{Adcox et~al.(2001)}]{Adcox:2000sp}
\bibinfo{author}{K.~Adcox}, et~al., \bibinfo{journal}{Phys.~Rev.~Lett.}
  \bibinfo{volume}{86} (\bibinfo{year}{2001}) \bibinfo{pages}{3500--3505}.
  \URLprefix \url{http://link.aps.org/doi/10.1103/PhysRevLett.86.3500}.
%Type = Article
\bibitem[{Aamodt and {ALICE Collab.}(2011)}]{ALICEPRL106}
\bibinfo{author}{K.~Aamodt}, \bibinfo{author}{{ALICE Collab.}},
  \bibinfo{journal}{Phys.~Rev.~Lett.} \bibinfo{volume}{106}
  (\bibinfo{year}{2011}) \bibinfo{pages}{032301}. \URLprefix
  \url{http://link.aps.org/doi/10.1103/PhysRevLett.106.032301}.
%Type = Article
\bibitem[{Eremin and Voloshin(2003)}]{EreminVoloshinPRC67}
\bibinfo{author}{S.~Eremin}, \bibinfo{author}{S.~Voloshin},
  \bibinfo{journal}{Phys.~Rev.~C} \bibinfo{volume}{67} (\bibinfo{year}{2003})
  \bibinfo{pages}{064905}. \URLprefix
  \url{http://link.aps.org/doi/10.1103/PhysRevC.67.064905}.
%Type = Article
\bibitem[{Nouicer(2007)}]{NouicerEPJC49}
\bibinfo{author}{R.~Nouicer}, \bibinfo{journal}{Eur.~Phys.~J.~C}
  \bibinfo{volume}{49} (\bibinfo{year}{2007}) \bibinfo{pages}{281--286}.
  \URLprefix \url{http://dx.doi.org/10.1140/epjc/s10052-006-0128-z}.
%Type = Article
\bibitem[{De and Bhattacharyya(2005)}]{DeBhattPRC71}
\bibinfo{author}{B.~De}, \bibinfo{author}{S.~Bhattacharyya},
  \bibinfo{journal}{Phys.~Rev.~C} \bibinfo{volume}{71} (\bibinfo{year}{2005})
  \bibinfo{pages}{024903}. \URLprefix
  \url{http://link.aps.org/doi/10.1103/PhysRevC.71.024903}.
%Type = Article
\bibitem[{Bialas(2008)}]{Bialas2008}
\bibinfo{author}{A.~Bialas}, \bibinfo{journal}{J. Phys.} \bibinfo{volume}{G 35}
  (\bibinfo{year}{2008}) \bibinfo{pages}{044053}. \URLprefix
  \url{http://dx.doi.org/10.1088/0954-3899/35/4/044053}.
%Type = Article
\bibitem[{Alver and Roland(2010)}]{AlverRolandPRC81}
\bibinfo{author}{B.~Alver}, \bibinfo{author}{G.~Roland},
  \bibinfo{journal}{Phys.~Rev.~C} \bibinfo{volume}{81} (\bibinfo{year}{2010})
  \bibinfo{pages}{054905}. \URLprefix
  \url{http://link.aps.org/doi/10.1103/PhysRevC.81.054905}.
%Type = Article
\bibitem[{Takahashi et~al.(2009)}]{TakahashiPRL103}
\bibinfo{author}{J.~Takahashi}, et~al., \bibinfo{journal}{Phys.~Rev.~Lett.}
  \bibinfo{volume}{103} (\bibinfo{year}{2009}) \bibinfo{pages}{242301}.
  \URLprefix \url{http://link.aps.org/doi/10.1103/PhysRevLett.103.242301}.
%Type = Article
\bibitem[{Tannenbaum(2006)}]{MJTCFRNC06}
\bibinfo{author}{M.~J. Tannenbaum}, \bibinfo{journal}{PoS}
  \bibinfo{volume}{(CFRNC2006)} (\bibinfo{year}{2006}) \bibinfo{pages}{001}.
  \URLprefix \url{http://pos.sissa.it/cgi-bin/reader/conf.cgi?confid=30}.
%Type = Article
\bibitem[{Alver et~al.(2006)}]{AlverCFRNC06}
\bibinfo{author}{B.~Alver}, et~al., \bibinfo{journal}{PoS}
  \bibinfo{volume}{(CFRNC2006)} (\bibinfo{year}{2006}) \bibinfo{pages}{023}.
  \URLprefix \url{http://pos.sissa.it/cgi-bin/reader/conf.cgi?confid=30}.

\end{thebibliography}






% \end{thebibliography}

\end{document}